\documentclass[10pt]{article}

\usepackage{amsmath}
\usepackage{amssymb}

\usepackage{graphicx}
\usepackage{psfrag}
\usepackage{subfigure}
\usepackage{cite}
\usepackage{color}
\usepackage{epic,eepic,rotating}

\usepackage{setspace} 
\usepackage[labelfont=bf,labelsep=period,justification=raggedright]{caption}

\makeatletter
\renewcommand{\@biblabel}[1]{\quad#1.}
\makeatother

\date{}

\begin{document}

\begin{flushleft}
{\Large
\textbf{Solving the Cable Equation Using a Compact Difference Scheme - Passive Soma Dendrite}
}
\\
Asha Gopinathan 
\\
\bf Department of Neurology, Sree Chitra Tirunal Institute of Medical Sciences and Technology,Tiruvananthapuram,Kerala,India 
\\
 E-mail: dendron.15@gmail.com
\end{flushleft}
\newpage
\section*{Abstract}
Dendrites are extensions to the neuronal cell body in the brain which are posited in several functions ranging from electrical and chemical compartmentalization to coincident detection. Dendrites vary across cell types but one common feature they share is a branched structure. The cable equation is a partial differential equation that describes the evolution of voltage in the dendrite. A solution to this equation is normally found using finite difference schemes. Spectral methods have also been used to solve this equation with better accuracy. Here we report the solution to the cable equation using a compact finite difference scheme which gives spectral like resolution and can be more easily used with modifications to the cable equation like nonlinearity, branching and other morphological transforms. Widely used in the study of turbulent flow and wave propagation, this is the first time it is being used to study conduction in the brain. Here we discuss its usage in a passive, soma dendrite construct. The superior resolving power of this scheme compared to the central difference scheme becomes apparent with increasing complexity of the model.
\section*{Keywords} dendrites, cable equation, passive soma-dendrite, numerical methods,spectral methods, finite difference scheme, compact difference scheme
\section*{Introduction}
\begin{figure}[!ht]
\begin{center}
\psfrag{a}{$a$}
\psfrag{ro}{$r_{o}$}
\psfrag{rm}{$r_{m}$}
\psfrag{cm}{$c_{m}$}
\psfrag{im}{$i_{m_{}}$}
\psfrag{ii}{$i_{i}$}
\psfrag{ri}{$r_{i}$}
\psfrag{X}{$X$}
\includegraphics[width = 3.27in]{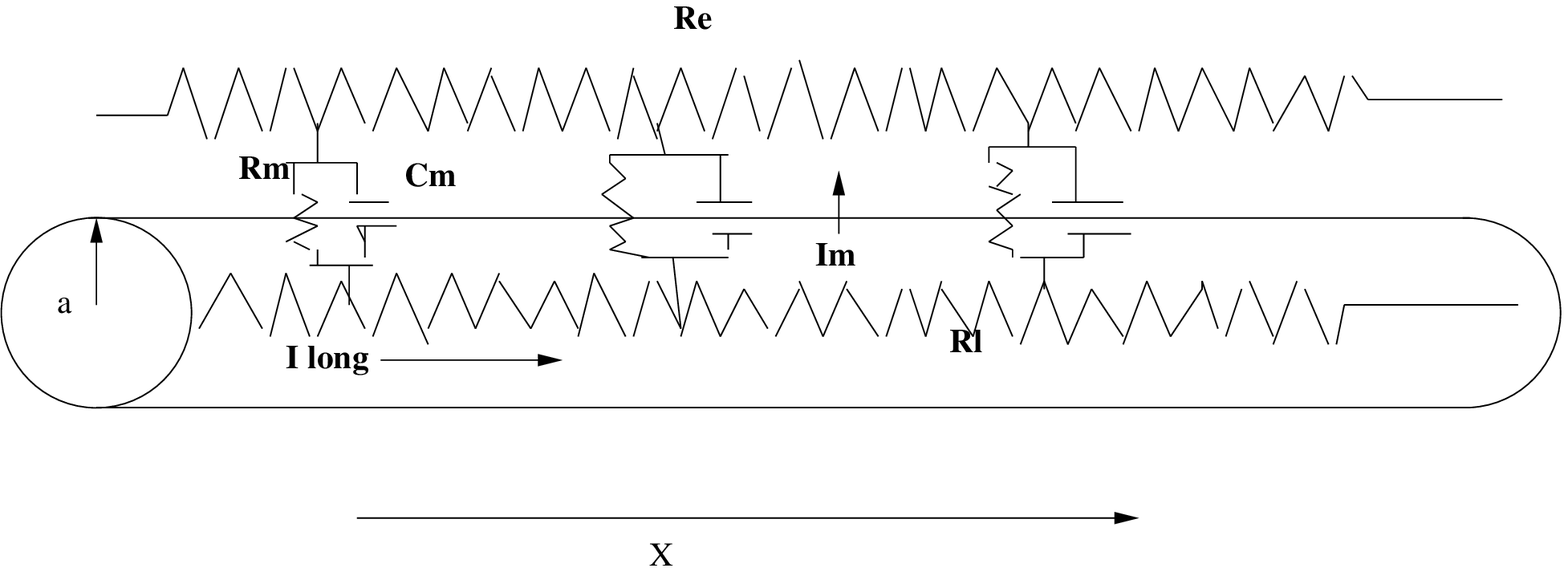}
\end{center}
\caption{\textbf{Compartmental electrical representation of a segment of passive cable}}
\label{fig:passivecompartmental}
\end{figure}

 The \textit{cable equation} governs the dynamics of membrane potential in thin and long cylinders such as axons or dendrites in neurons. It is Wilfred Rall's derivation of the cable equation which set the ground for describing quantitatively the change in voltage across space and time in dendrites ~\cite{rall:59,koch:99,tuckwell1:88a}. Under consideration here is a linear system involving neuronal processes that have a voltage independent component. The model is a \textit{passive} one which means that membrane conductance is fixed. Here we only look at \textit{subthreshold responses} which are about half the strength required to generate action potentials. If the response is stronger the membrane potential will begin to change and our assumption of \textit{passive dendrite} will no longer be true.  The cable equation is (~\cite{koch:99}, equation $2.7$)
\begin{equation}
\frac{\tau_{m}\partial V_{m}(x,t)}{\partial t} = \frac{\lambda^{2} \partial^{2}V_{m}(x,t)}{\partial x^{2}} - (V_{m}(x,t)- V_{rest}) + r_{m}I_{inj}(x,t)
\end{equation} 
Here $V_{m}$ is the membrane potential in mV, $r_{m}$ is the membrane resistance per unit length of the fibre in $\Omega$.cm. $I_{inj}(x,t)$ is the injected current in Amperes, $\tau_{m}= r_{m}c_{m}$ is the membrane time constant and $\lambda = (r_{m}/r_{a})^{1/2}$ is the membrane space constant. $c_{m}$ is capacitance per unit length of cable of diameter $ d $ in units of F/cm. $r_{a}$ is axial resistance per unit length in $\Omega$/cm. $V_{rest}$ varies between $-50$ and $-90$ mV depending on cell type.  
Nondimensionalising this equation results in :
\begin{equation}
 \frac{\partial V(X,T)}{\partial T} = \frac {\partial^{2}V (X,T)}{\partial X^{2}} - V(X,T) + \frac{I(X,T)}{\lambda c_{m}}
\end{equation} 
Here $ X= x/\lambda $, $T= t\tau $, $I(X,T)=\lambda \tau I_{inj}(x,t)$, $I_{inj}(x,t)$ is stimulus current density and $c_{m}=C_{m}\pi d$, $C_{m}$ is specific capacitance/unit area in F/ $cm^{2}$. Fig~\ref{fig:passivecompartmental} shows the compartmental electrical representation of a segment of passive membrane.
\section*{\begin {normalsize}\textbf{Numerical method for the cable equation}\end{normalsize}}
\begin{figure}[!ht]
\begin{center}
\psfrag{point soma}{$point soma$}
\psfrag{node}{$node$}
\psfrag{dendrite}{$dendrite$}
\psfrag{sealed/killed end}{$sealed/killed end$}
\psfrag{h}{$h$}
\psfrag{l}{$l$}
\includegraphics[width = 3.27in]{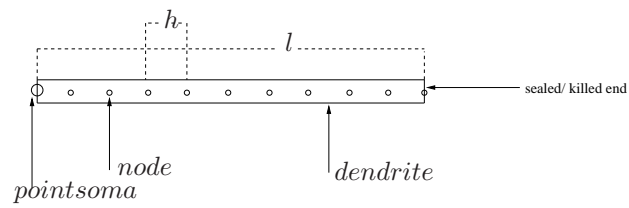}
\end{center}
\caption{\textbf{Discretisation of a dendritic tree}}
\label{fig:pointdiscrete}
\end{figure}

Numerical analysis involves designing algorithms that approximate the solution to the equation under consideration~\cite{moin:01}. As in all approximations, this results in a small error called the truncation error. Most finite difference schemes use the central difference formulae. It can be derived from the Taylor's series and shows algebraic convergence~\cite{lindsay:99}. The second derivative $\frac{\partial^{2}V}{\partial X^{2}}$ in equation 1, also written as $V''$ can be approximated as (~\cite{lindsay:99},equation 8.54) :
\begin{equation}
V_{i}'' = \frac{(V_{i+1}-2V_{i}+V_{i-1})}{h^{2}} + O(h^{2})  ,  2\leq i \leq N-1
\end{equation} 
The subscripts denote locations on a uniformly spaced grid $x_{i}$ and h is the mesh size $x_{i+1}-x_{i}$. $O(h^{2})$ is the truncation error which is second order in this case. The length $L$ of the cable is divided into uniform segments of length $ h= L/(N-1)$ where $ N $ is the number of nodes. $V_{i}$ is the voltage at location $i$ at distance $ X_{i} = (i-1)h$. Indices $ i = 1$ and $ i = N$ denote boundary points $X=0$ and $X= L$. Fig~\ref{fig:pointdiscrete} shows this. 
\par\parindent = 1in For spectral methods the convergence is exponentially fast~\cite{lindsay2:01}. The cable equation has been solved using a spectral Chebyshev method~\cite{toth:99}. The solutions have uniform, high numerical accuracy at \textit{any} spatial point- not just the original collocation points. The truncation error is negligible and the total error can be the roundoff error of the computations. 
\newline In this paper a compact finite difference scheme is utilized~\cite{lele:92}. Such a scheme provides better resolution at the shorter length scales compared to the usual explicit schemes like equation $3$. Thus it provides the resolution of spectral methods but unlike spectral methods which are sensitive to discontinuities, requires little extra procedure for generalisation of the governing equation ( like geometry and other property variations along the dendrite). 
\section*{\begin{normalsize}\textbf{Spatial discretisation : Using compact finite difference schemes to solve the cable equation}\end{normalsize}}
The second derivative $\frac{\partial^{2}V}{\partial x^{2}}$ is approximated using the following equation (~\cite{lele:92}, equation $2.2$):
\begin{eqnarray}
\beta V''_{i-2} + \alpha V''_{i-1} + V''_{i} + \alpha V''_{i+1} + \beta V''_{i+2}& = & \frac{c (V_{i+3}-2V_{i}+V_{i-3})}{9h^{2}}\nonumber \\ & &+\frac{b(V_{i+2}-2V_{i}+V_{i-2})}{4h^{2}} \nonumber \\
& & +\frac{ a( V_{i+1}-2V_{i}+V_{i-1})}{h^{2}}   ,  2 \leq i \leq N -1
\end{eqnarray} 
where $V''_{i}$ represents the finite difference approximation to the second derivative at node i.
 The relations between the coefficients a,b,c and $\alpha$ , $\beta$ are derived by matching the Taylor series coefficients of various orders. 
We take (~\cite{lele:92},equation $2.2.7$)
\begin{displaymath}
\alpha = \frac{2}{11}, \beta = 0, a = \frac{12}{11}, b =\frac{ 3}{11}, c= 0
\end{displaymath}
to obtain a sixth order formula. For the boundaries the scheme chosen is (~\cite{lele:92},equation $4.3.4$)
\begin{equation}
V''_{1}+ \alpha V''_{2} = \frac{ aV_{1}+bV_{2}+cV_{3}+dV_{4}+eV_{5}}{h^{2}}
\end{equation} 
A similar equation connects $V''_{N}$ and $ V''_{N-1}$. 
By requiring third - order formal accuracy the coefficients are reduced to (~\cite{lele:92},equation $4.3.6$),choosing $\alpha=\frac{1}{10}$ from classical Pade scheme which is fourth order.
\begin{eqnarray*} 
a = \frac{(11 \alpha + 35)}{12}, \hspace{2mm} b = \frac{- (5\alpha +26)}{3} ,\hspace{2mm} c = \frac{(\alpha +19)}{2}, \hspace{2mm} d =\frac{(\alpha -14)}{3}\\  e = \frac{(11 - \alpha)}{12}
\end{eqnarray*}
The truncation error is reduced to $ (\frac{\alpha -10}{12})h^{3}V^{5}$. If $\alpha$ is $10$, truncation error becomes $h^{4}$. Equation $4$ and $5$ applied at interior points result in a matrix problem $\textbf{A}V'' = \textbf{B}$ where A is tridiagonal and $V''$ can be obtained easily.
\section*{\begin{normalsize}\textbf{Time discretisation}\end{normalsize}}
\par\parindent = 1in The values for $V''$ calculated from the compact-difference scheme were used to integrate the result in time using an explicit time stepping scheme - forward Euler. If $ t = n\Delta T$, $V^{n} \equiv V(t)$ and $V^{n+1} \equiv V(t+ \Delta T)$, then :
\begin{equation}
 V^{n+1}= V^{n} + f(V^{n},n\Delta T)\Delta T
\end{equation} 
 Stability conditions have dictated the choice of the time step
 \begin{equation}
  \Delta T = \frac{\Delta x^{2}}{4}                                                                                                                                                                                                                                                                                                                                                                                                                                                                                                                                                               
  \end{equation} 
It varies as shown in (Table~\ref{tab:DeltaT}).The numerical integration in time has been done with an explicit scheme. Since spatial derivatives are obtained with a compact scheme, which is an implicit formula that requires the solution of a linear system, implicit time-stepping is not possible. Implicit time-stepping is desirable to overcome the severe restrictions that stability imposes on the time-step of conditionally stable explicit schemes. A work-around is to use a predictor-corrector scheme which uses an explicit step estimate from the predictor step in a corrector step which is also an explicit step.\\
Computations were performed on a Toshiba Satellite Pro laptop using Octave in a Linux(Ubuntu)environment. The data used for  simulations is given in (Table~\ref{tab:parameters}).  
\section*{\begin{normalsize}\textbf{Specific problems}\end{normalsize}}
\par
\begin{figure}[!ht]
\begin{center}
\psfrag{Soma}{$Soma$}
\psfrag{Dendrite}{$Dendrite$}
\psfrag{Current Injection}{$Current Injection$}
\psfrag{d}{$d$}
\psfrag{l}{$l$}
\subfigure[\textbf{Point soma dendrite construct with current injection at $i = 1$}]{
\includegraphics[width=2.0 in]{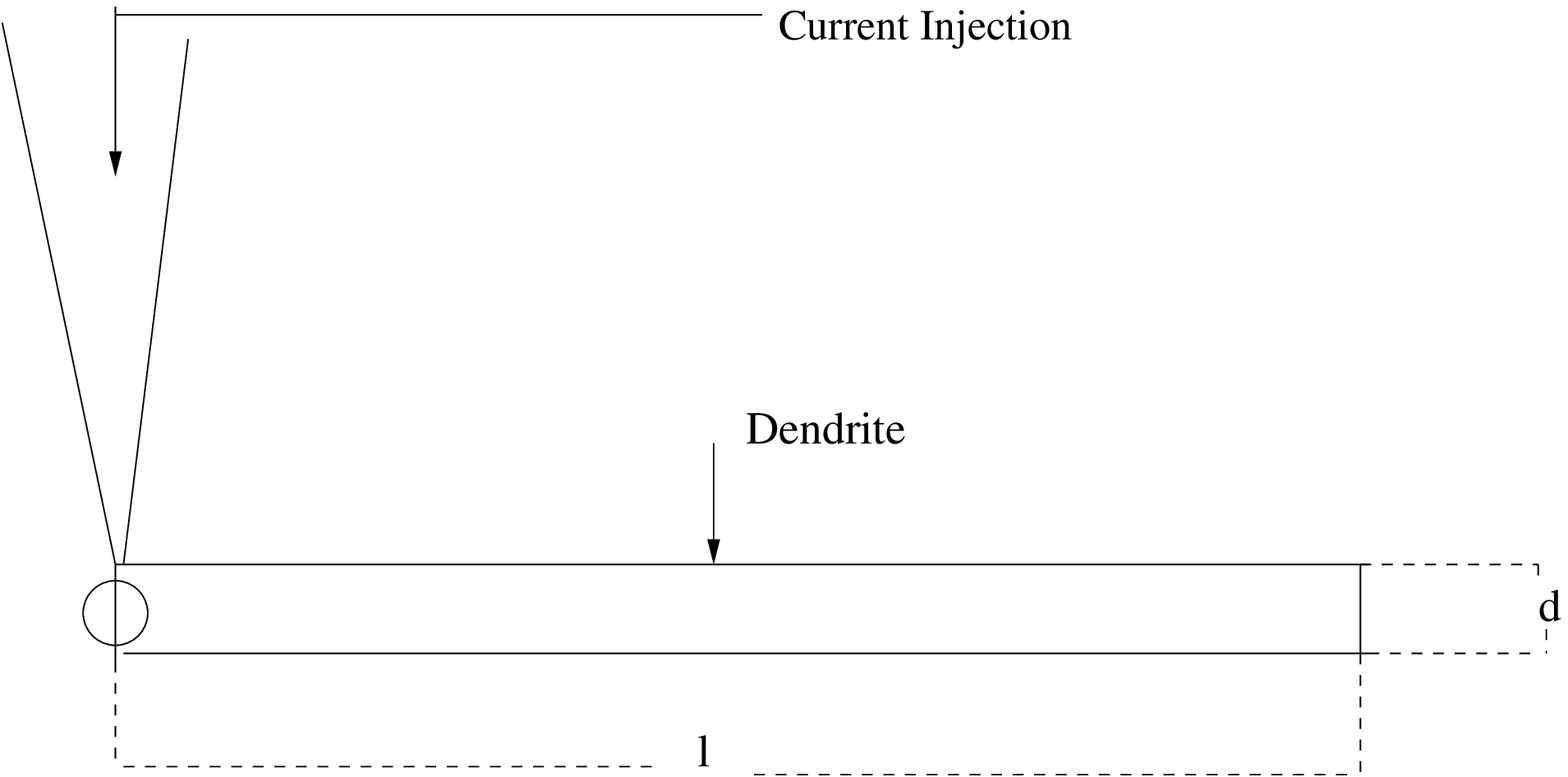}
}
\subfigure[\textbf{Soma dendrite construct with lumped soma boundary condition}]{
\includegraphics[width = 2.0 in]{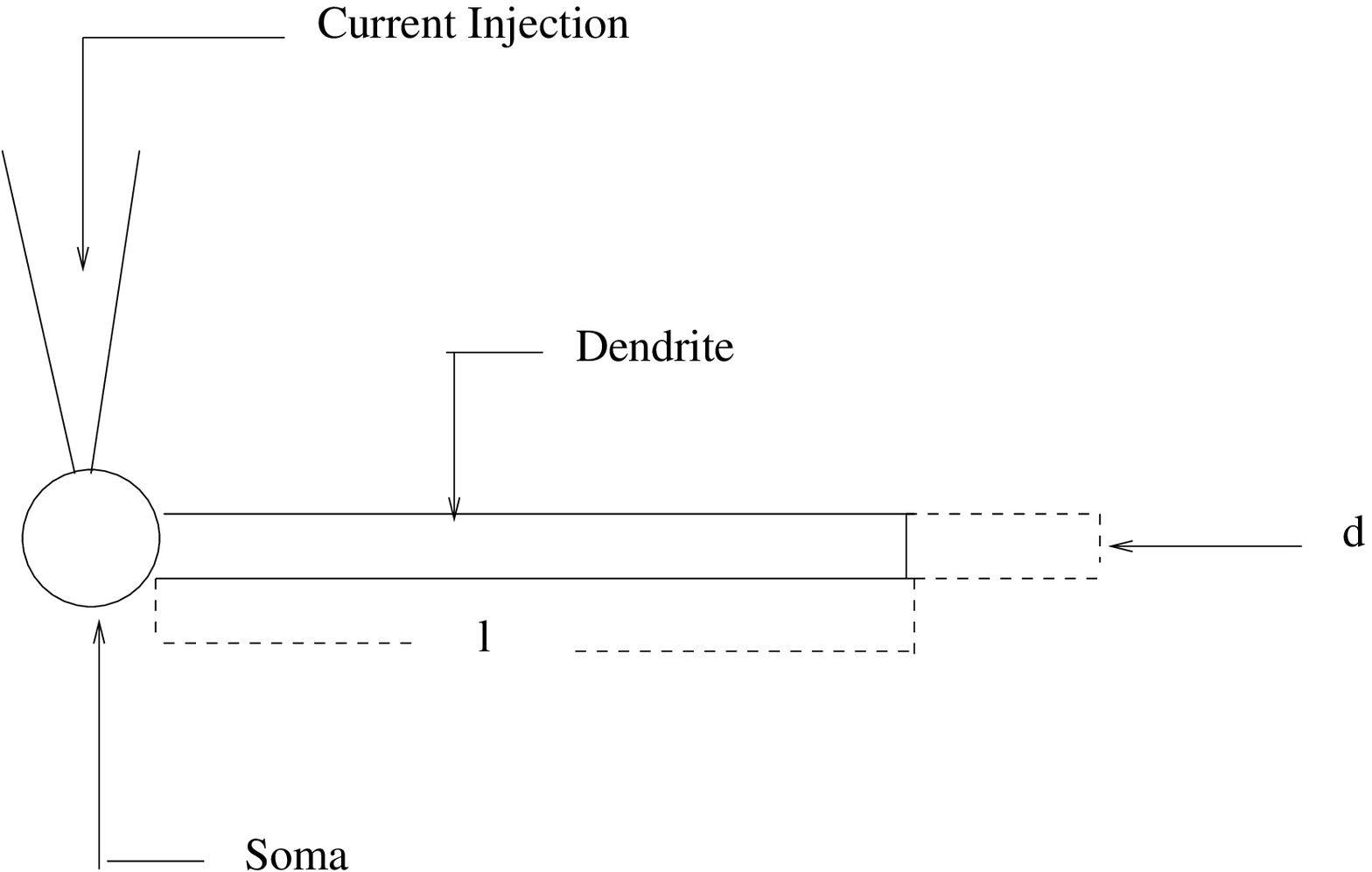}
}
\subfigure[\textbf{Point soma dendrite construct with current injected at $ i= N$}]{
\includegraphics[width = 2.0 in]{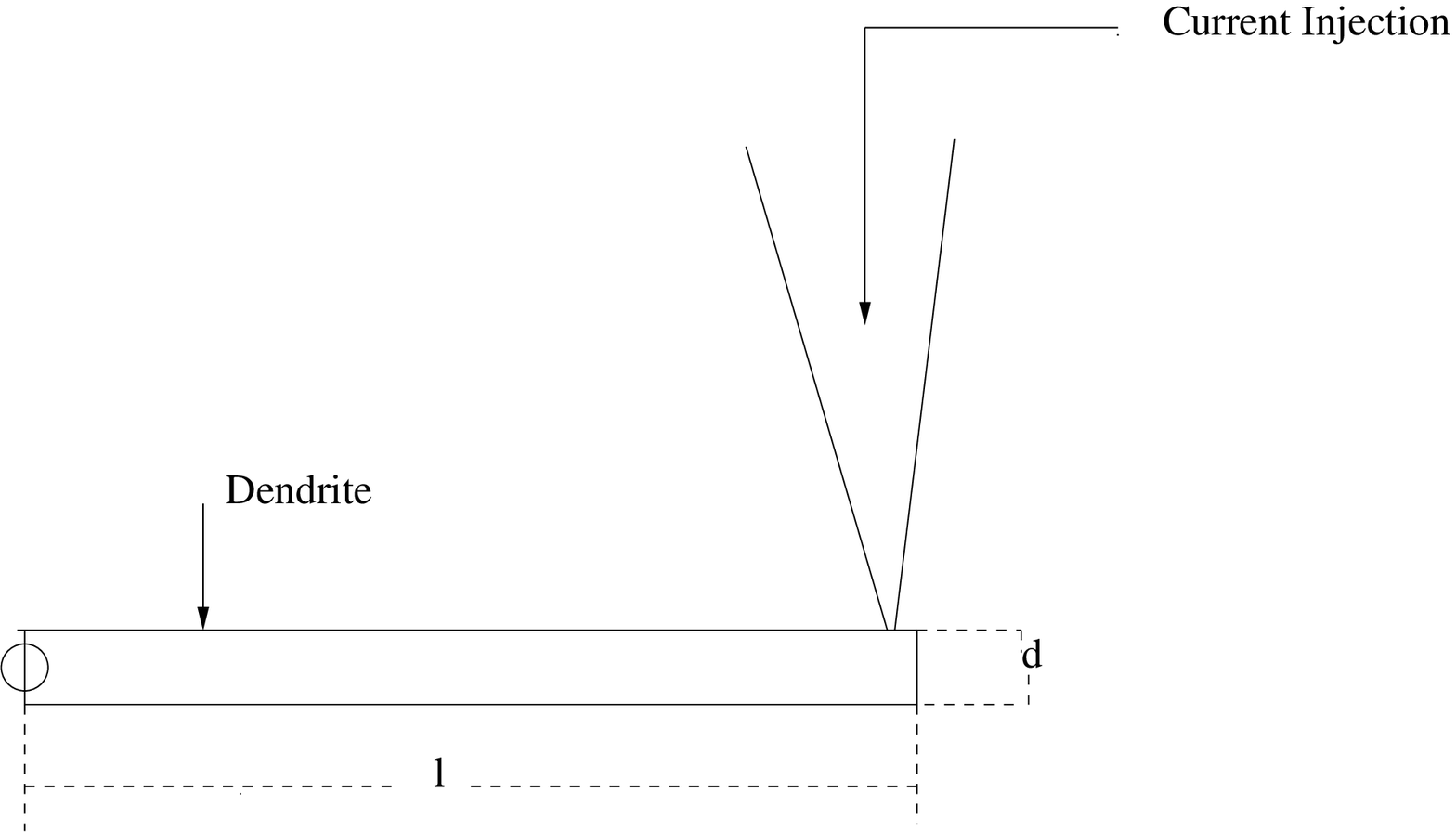}
}
\subfigure[\textbf{Point soma dendrite construct with current injected at $ i= N/2$}]{
\includegraphics[width = 2.0 in]{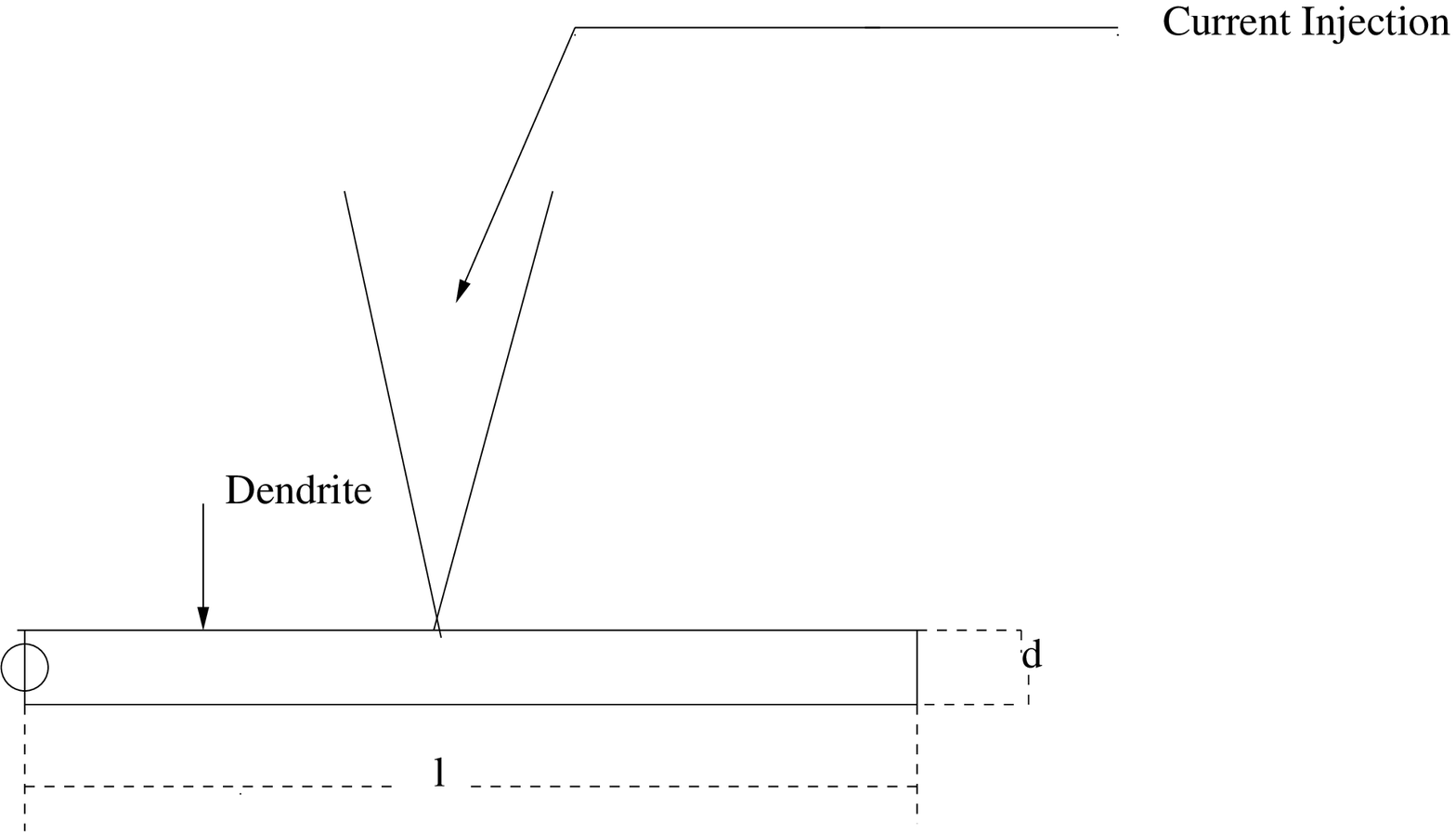}
}
\end{center}
\caption{\textbf{Soma dendrite construct - different cases}}
\label{fig:pointsomadendx0}
\end{figure}

\parindent = 1in There are four problems we have considered. Before discussing each in detail, the initial and boundary conditions used are discussed:\\
\textbf{Initial Condition}:
\begin{equation}
 V(x,0) = v(x) , 0 \leq x \leq l
\end{equation} 
Here $v(x) = -70$ mV. 
\newline
\textbf{Boundary Condition}:
\newline
Notation : $V'(x,t)= \frac{\partial V(x,t)}{\partial x}$
\newline
\textbf{Current Injection}:
\newline
If there is a current of magnitude $I(t)$ injected into the end at $x=0$ ( in the positive x- direction),(~\cite{tuckwell1:88a},equation $4.36$)
\begin{equation}
 V'(0,t) = -r_{i}I(t), t>0
\end{equation} 
If the current $I(t)$ is injected into the end at $x=l$,(~\cite{tuckwell1:88a},equation $4.37$)
\begin{equation}
 V'(l,t) = r_{i}I(t), t>0
\end{equation} 
\newline
\textbf{Sealed end}
\newline
If end at $x=0$ is sealed,(~\cite{tuckwell1:88a},equation $4.32$)
\begin{equation}
 V'(0,t)=0, t>0
\end{equation} 
If end at $x=l$ is sealed,(~\cite{tuckwell1:88a},equation $4.33$) :
\begin{equation}
 V'(l,t) = 0,t>0
\end{equation} 
\textbf{Killed end}
\newline
If end at $x=0$ is killed,(~\cite{tuckwell1:88a},equation $4.34$)
\begin{equation}
 V(0,t)=0,t>0
\end{equation} 
or if end at $x=l$ is killed,(~\cite{tuckwell1:88a},equation $4.35$)
\begin{equation}
 V(l,t)=0,t>0
\end{equation} 
\textbf{1.Point soma dendrite, current injection at $x = 0$}: The point soma dendrite construct we have used is a soma( cell body)collapsed into a point as mentioned in ~\cite{toth:99} along with a dendrite attached to it (Fig.~\ref{fig:pointsomadendx0}). The governing equation is the cable equation as mentioned above in equation $1$. (  Range for independent variables $ 0 <x< l, t > 0 $ and the steady state at $ t \rightarrow \infty) $. Initial condition is given by equation $8$. Boundary conditions are current injection at $i=1$ (equation $9$) and at $i=N$, sealed end (equation $12$) or killed end(equation $14$). \newline The analytical equation at steady state for this formulation with sealed end boundary condition at $i=N$ is (~\cite{tuckwell1:88a}, Table $4.2$) :
\begin{equation}
 V(x) = V(0)(\frac{Cosh(L-X)}{Cosh(L)})
\end{equation} 
where $V(x)$ is the voltage at any given $i$, $V(0)$ is the voltage at $i=1$ , $L = \frac{l}{\lambda}$, $X= \frac{x}{\lambda}$. 
The analytical equation at steady state for the boundary condition with killed end at $i=N$  is (~\cite{tuckwell1:88a}, Table $4.2$):
\begin{equation}
 V(x) = V(0)(\frac{Sinh(L-X)}{Sinh(L)})
\end{equation} 
where $V(x)$, $V(0)$,$L$ and $X$ are as defined above. \\
\textbf{2. Lumped soma model,current injected into soma}: Additionally, we have used a lumped soma model where the current is injected into the soma (Fig.~\ref{fig:pointsomadendx0}). 
Here the boundary condition at $i=1$ is a lumped soma boundary condition at steady state ( ~\cite{tuckwell1:88a}, equation $ 6.14$),
For a first order case,this is : 
\begin{equation}
 \frac{R_{s}\bar{I_{s}}}{\tau} = \bar\tilde{V(0)}- \gamma \bar\tilde{V'(0)}
\end{equation} 
where $Rs$ is the membrane resistance at soma, $\gamma = \frac{R_{s}}{r_{i}}$, $I_{s}$ is the current injected at the soma, $\tau$ is the time constant. \newline The boundary condition at $i=N$ can either be sealed end ( equation $12$) or killed end ( equation $14$).
The analytical equation at steady state is (~\cite{tuckwell1:88a},equation $6.20$) : 
\begin{equation}
 V(X)= IR_{N}(\frac{ Cosh(L-X)}{Cosh(L)})
\end{equation} 
where $V(X)$ is the voltage at any given node $i$, $I$ is the injected current, $L= \frac{l}{\lambda}$, $X = \frac{x}{\lambda}$ and  $\frac{1}{R_{N}}= \frac{1}{R_{S}}+ \frac{1}{R_{M}}$ (~\cite{tuckwell1:88a},equation $6.21$).\\
\textbf{3. Point soma dendrite, current injection at $ i = N$}: When current is injected at the end of the dendrite at $i=N$ as shown in (Fig.~\ref{fig:pointsomadendx0}), then according to convention longitudinal currents are positive in the positive x- direction.
The equation governing this is the cable equation (1) as before with conditions given earlier. The boundary condition for current injected at the end ($i= N$) is defined by equation $10$. The boundary conditions at the end ($i=1$) can be sealed ( equation $11$) or killed ( equation $13$).\\
\textbf{4. Point soma dendrite, current injection at $ i = N/2$}: Here current is injected at the point $i=N/2$ as shown in (Fig.~\ref{fig:pointsomadendx0}) 
\newline The current injection condition is
 \begin{equation}
 I(X,T) = \lambda \tau I_{inj}(x,t) , i = N/2
\end{equation}
\begin{equation}
I(X,T) = 0, i \neq N/2
 \end{equation}
The boundary conditions at $i=1$ can be sealed ( equation $11$) or killed (equation $13$). Similarly the boundary conditions at $i=N$ can be sealed ( equation $12$) or killed ( equation $14$).
\newline The aim of this paper is to solve the cable equation using the compact difference scheme. This has been achieved in the passive soma dendrite construct. To the best of our knowledge, this is the first time the compact difference scheme is being used to solve equations involving changing parameters in the brain. 
                                                                                                                                                                                                                                                                                                                                                                                                                                                                                                                                                           
\section*{Results}
The results obtained in the above discussed simulations are presented here:
\subsection*{Point soma dendrite, current injection at $x=0$}
\subsubsection*{Current injection at $i=1$, Evolution of voltage}
\begin{figure}[!ht]
\begin{center}
\psfrag{V}{$V$}
\psfrag{X}{$X$}
\subfigure[{sealed end }]{
\includegraphics[width = 3.0in]{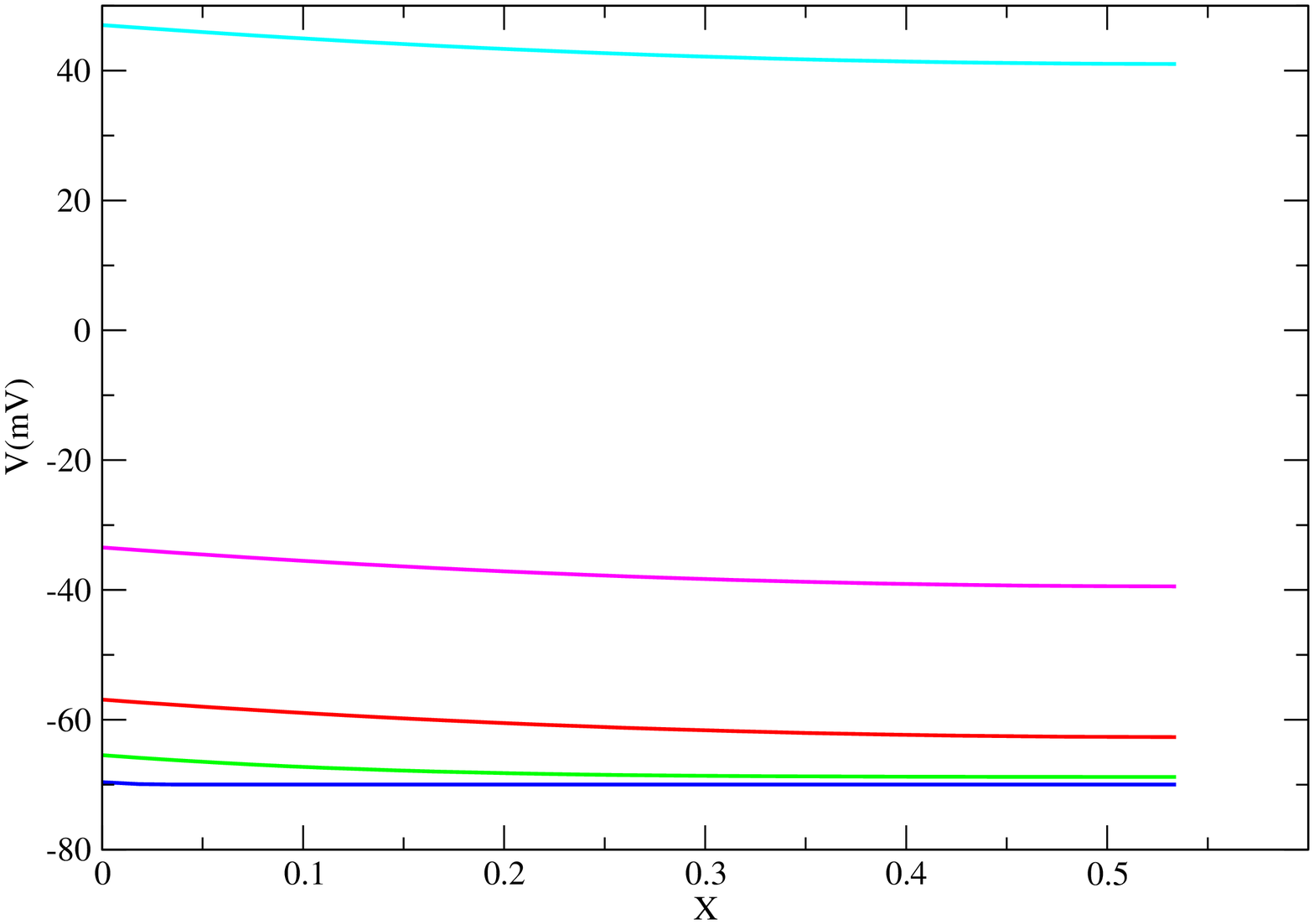}
}
\psfrag{V}{$V$}
\psfrag{X}{$X$}
\subfigure[{killed end }]{
\includegraphics[width= 3.0in]{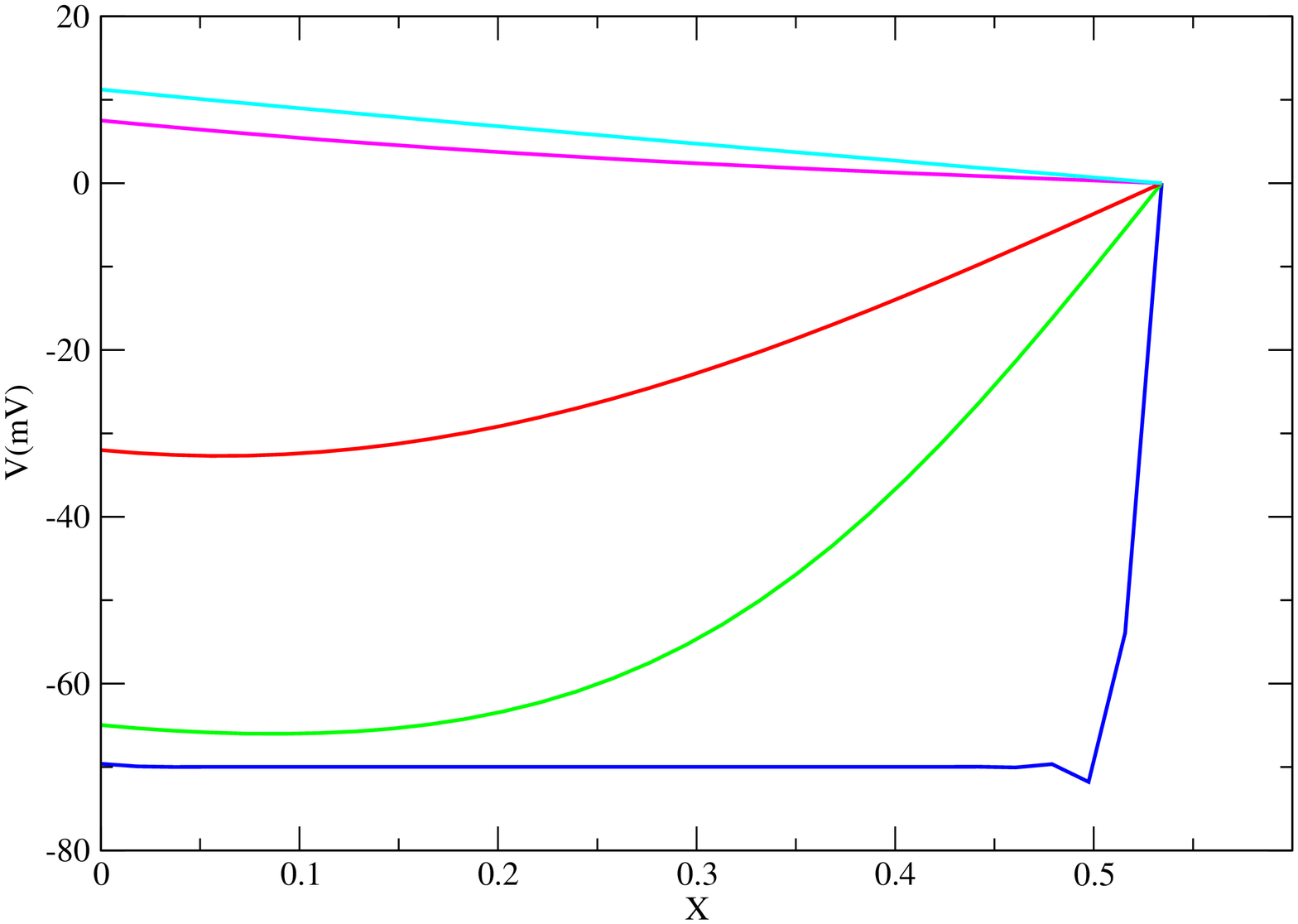}
}
\psfrag{V}{$V$}
\psfrag{X}{$X$}
\subfigure[{sealed end,lumped soma }]{
\includegraphics[width = 3.0in]{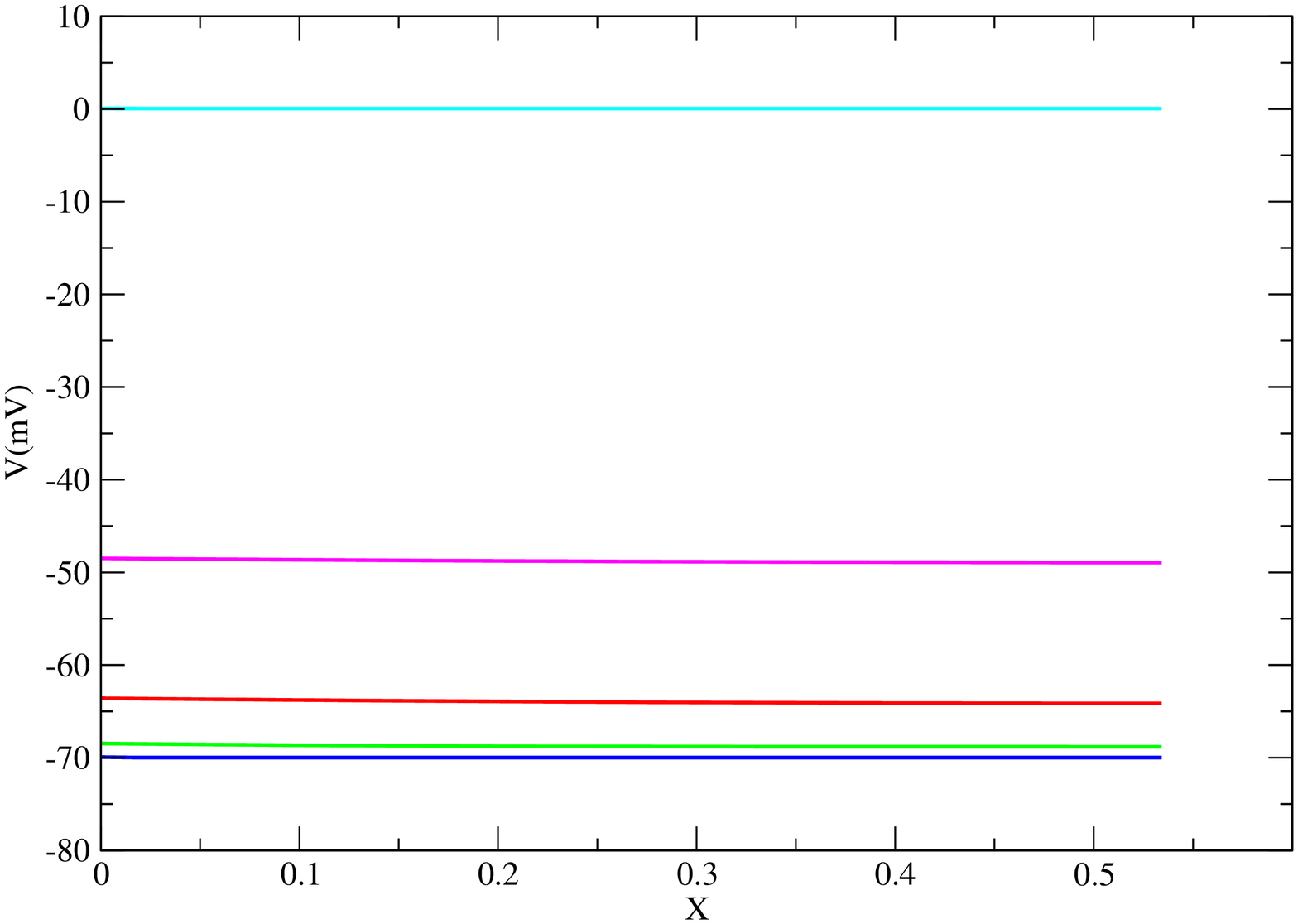}
}
\psfrag{killed end N=30}{$killed end N=30$}
\psfrag{V}{$V$}
\psfrag{X}{$X$}
\subfigure[{killed end,lumped soma }]{
\includegraphics[width= 3.0in]{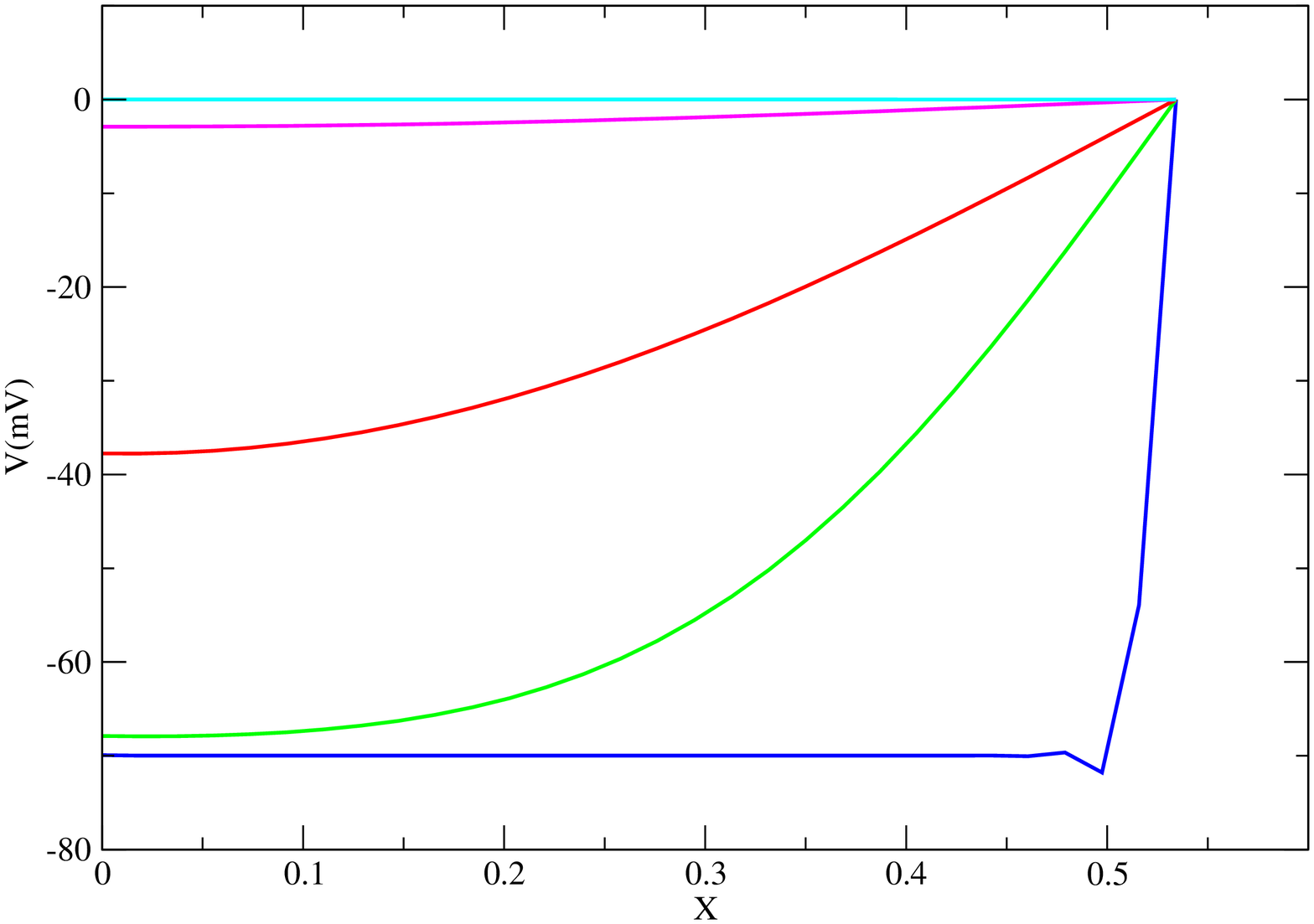}
}
\end{center}
\caption{\textbf{Evolution of voltage point and lumped soma, current injected at $i=1$,{\color{blue} ------}: $n=1$; {\color{green} ------}: $n=100$;
{\color{red}-----}:$n==500$;{\color{magenta}-----}:$n==2000$;{\color{cyan}----}:$n==294670$} }
\label{fig:sevoltNoct2}
\end{figure}

In this simulation, current is injected at $i=1$ and the evolution of voltage at various times is observed.\\
\begin{equation}
 t= (\Delta T)(n)(\tau) 
\end{equation}
For $N=30$,$\tau = 20$ msec,the voltage evolution is looked  at  $t$ as shown in (Table~\ref{tab:niter}). The results for both sealed and killed end boundary condition at $i=N$ can be seen in (Fig~\ref{fig:sevoltNoct2}). 
\subsubsection*{Current injection at $i=1$, Comparison with analytical equation at steady state}
\begin{figure}[!ht]
\begin{center}
\psfrag{sealed end N=30,V/Vo vs X}{$sealed end N =30,V/Vo vs X$}
\psfrag{V/Vo}{$V/Vo$}
\psfrag{V/Vo1}{$V/Vo$}
\psfrag{X(lambda)}{$X(lambda)$}
\subfigure[{sealed end}]{
\includegraphics[width = 3.0in]{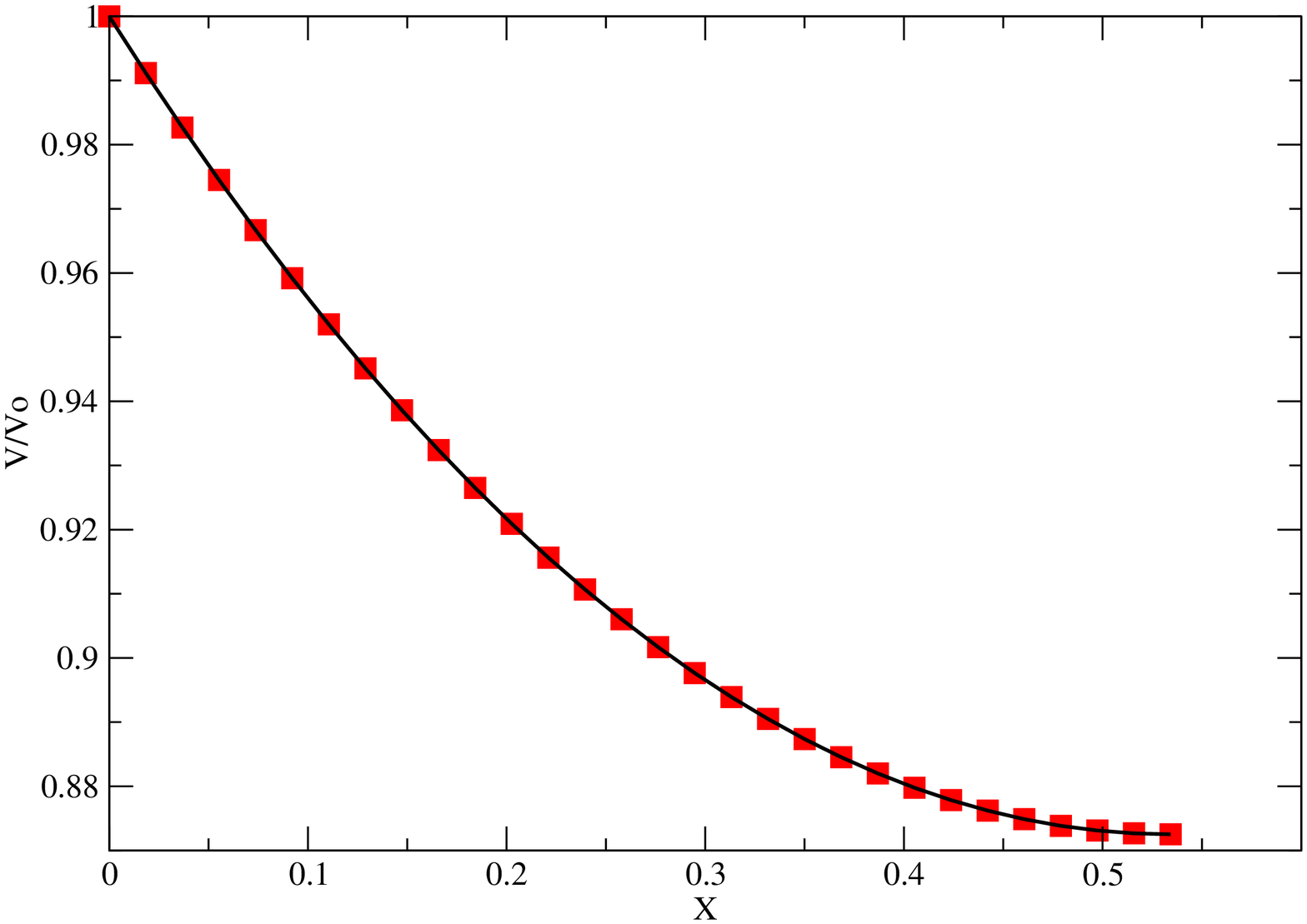}
}
\psfrag{killed end N=30,V/Vo vs X}{$killed end N =30,V/Vo vs X$}
\psfrag{V/Vo}{$V/V0$}
\psfrag{X(lambda)}{$X(lambda)$}
\subfigure[{killed end}]{
\includegraphics[width =3.0in]{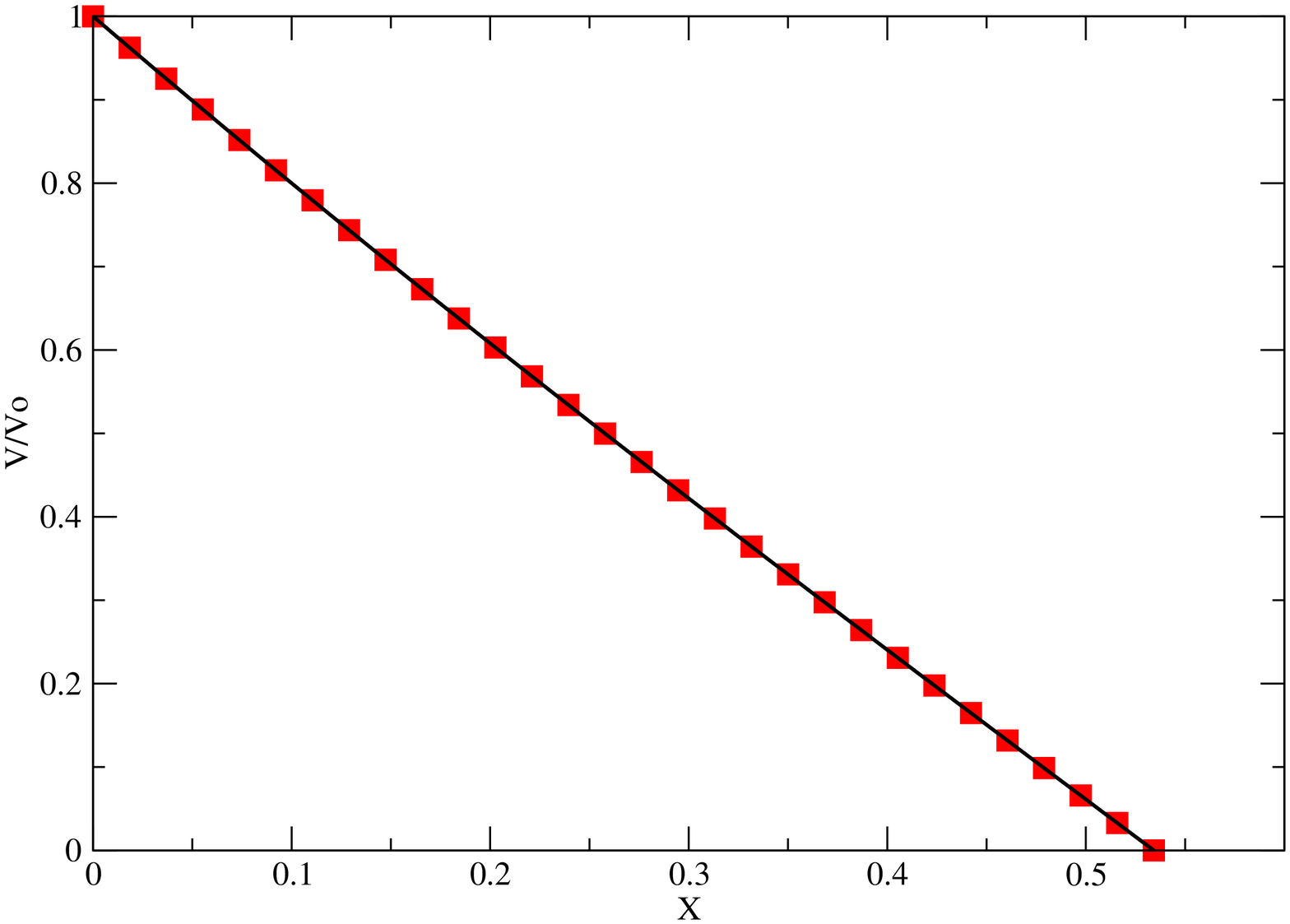}
}
\end{center}
\caption{\textbf{Comparison of compact scheme with analytical solution, sealed and killed end,point soma  { \color{red}$\blacksquare$}:compact;{\color{black}----}:analytical}}
\label{fig:sevoltanaly}
\end{figure}

In this simulation, current is injected at $i=1$, $N=30$ and the simulation is run till $ t = 500$ msec. The results are compared with the analytical equation. $V/V_{0}$ is plotted against $X$. The results are shown in (Fig.~\ref{fig:sevoltanaly}).
\subsubsection*{Current injection at $i=1$, Comparison between compact, central difference schemes with analytical equation at steady state, sealed end boundary condition}
\begin{figure}[!ht]
\begin{center}
\psfrag{V/Vo}{$V/Vo$}
\psfrag{X(lambda)}{$X(lambda)$}
\psfrag{a}{$a$}
\psfrag{b}{$b$}
\psfrag{c}{$c$}
\psfrag{d}{$d$}
\subfigure[{$N=10$}]{
\includegraphics[width = 3.0in]{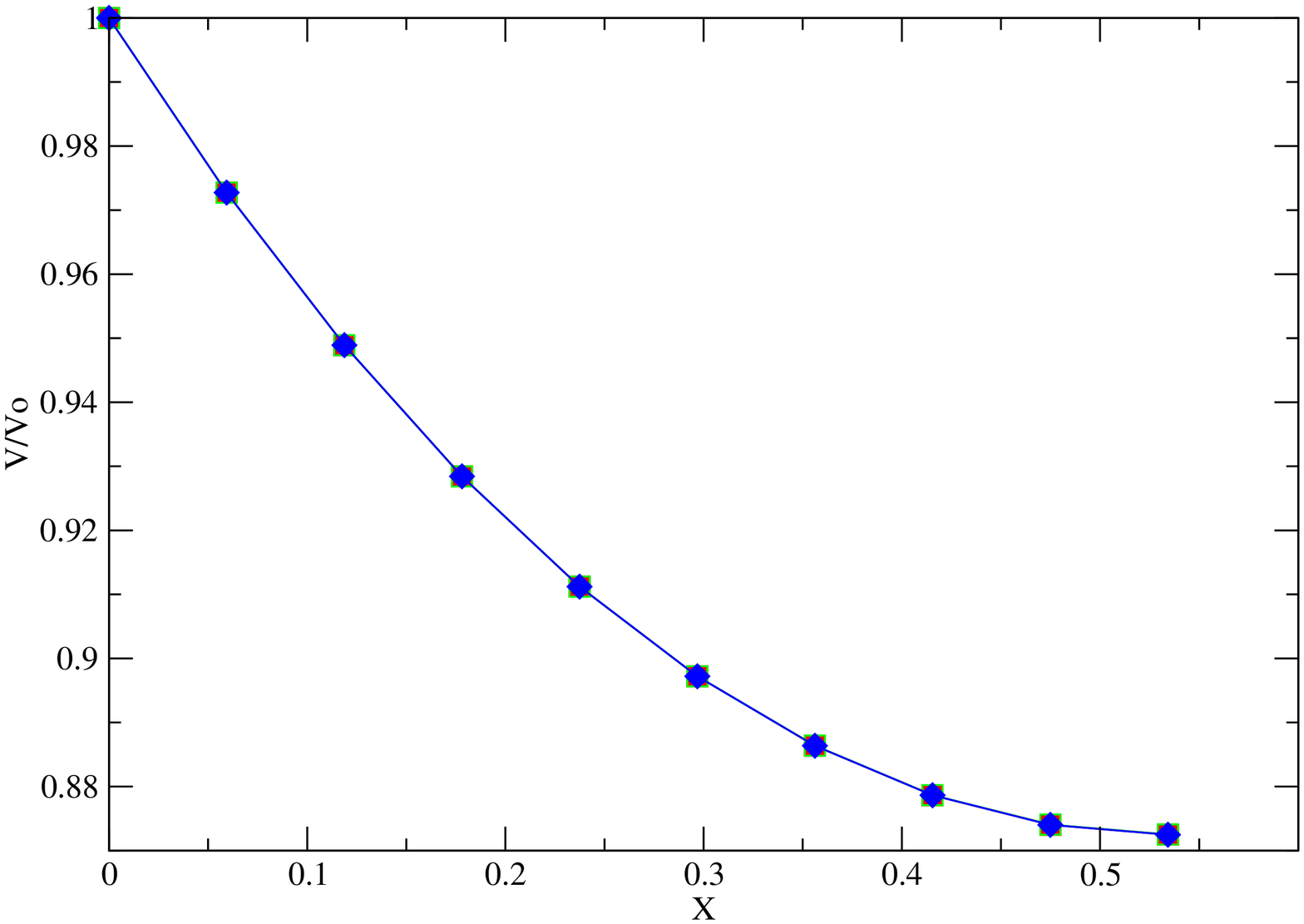}
}
\subfigure[{$N=20$}]{
\includegraphics[width = 3.0in]{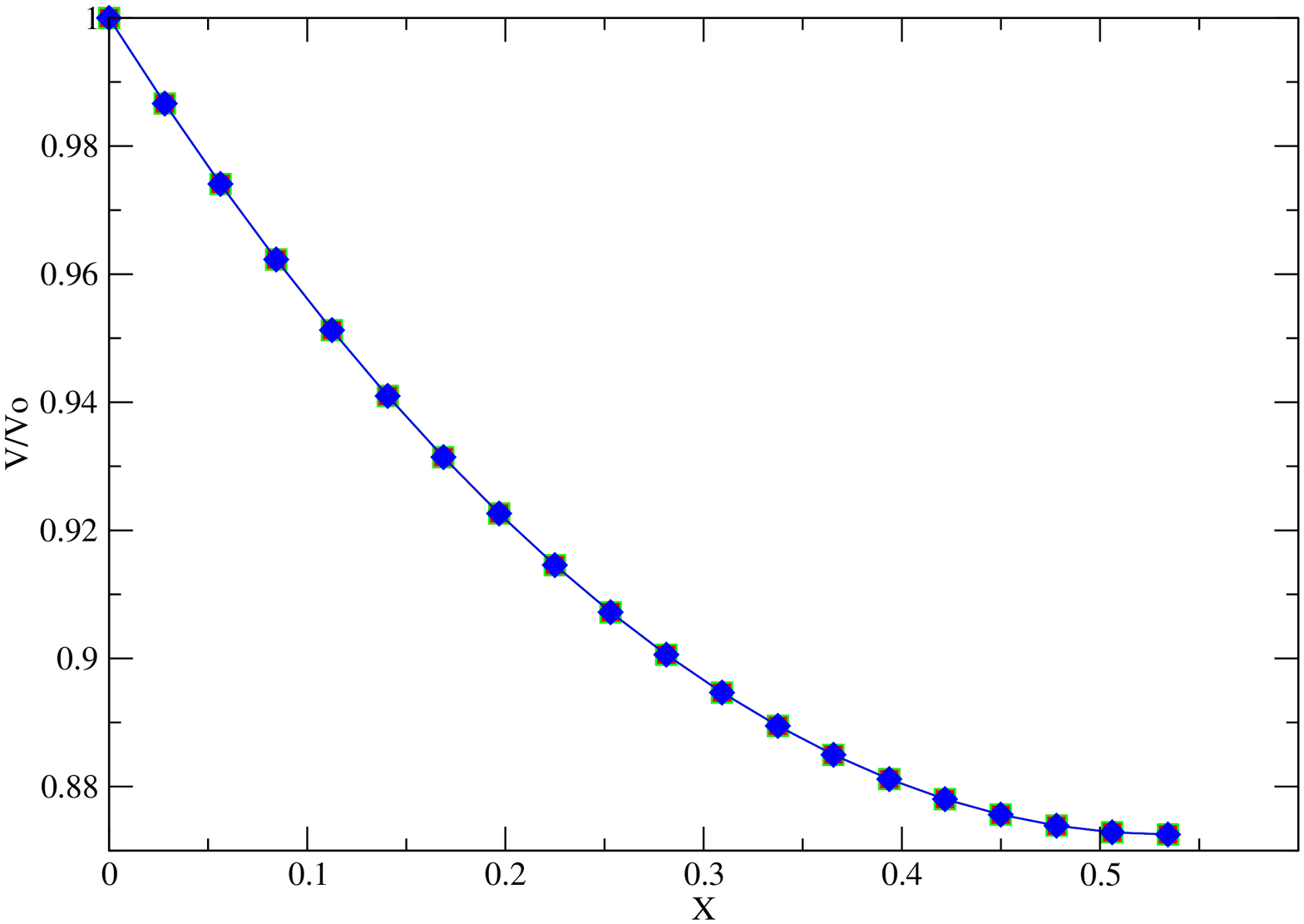}
}
\subfigure[{$N=30$}]{
\includegraphics[width = 3.0in]{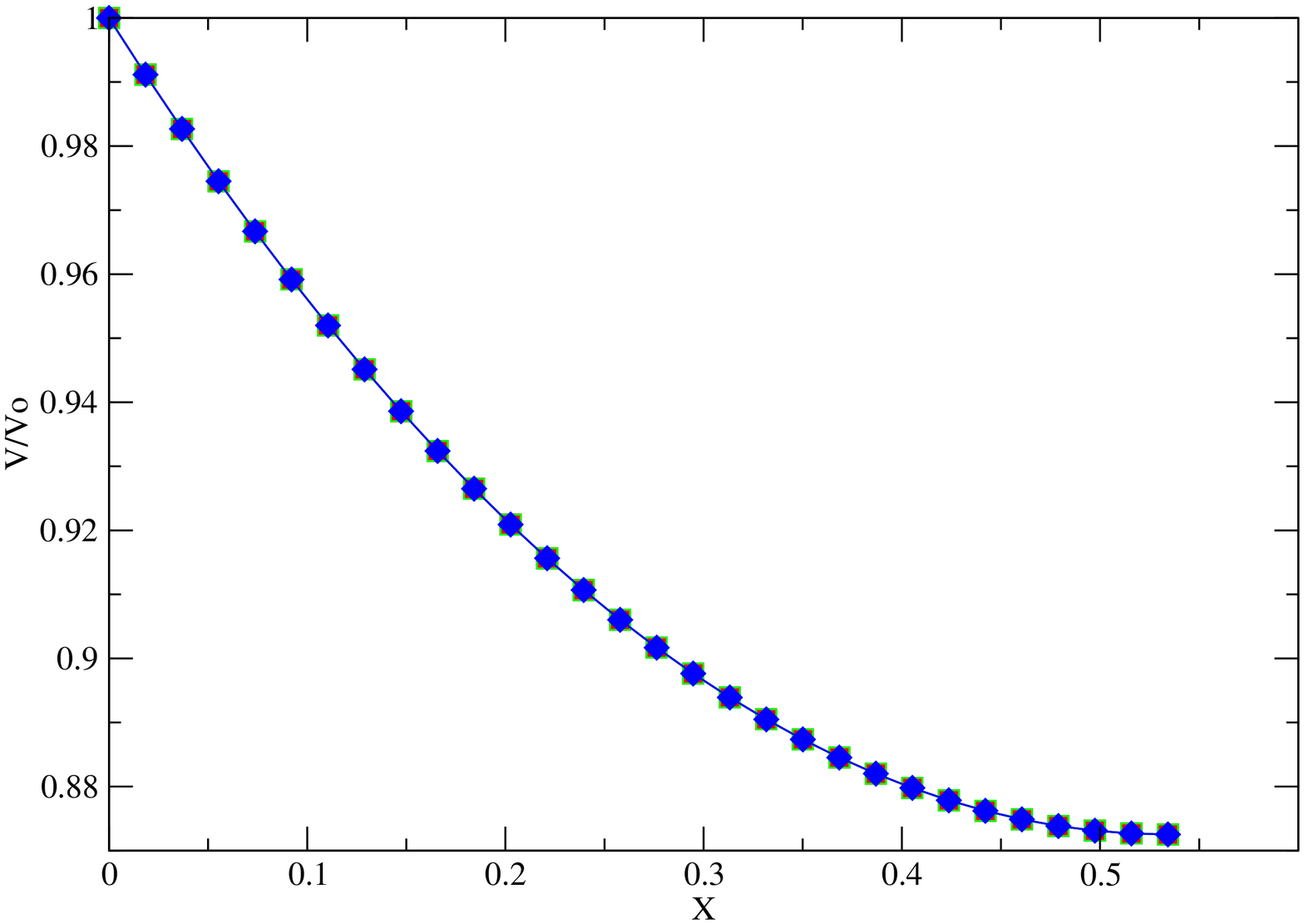}
}
\subfigure[{$N=40$}]{
\includegraphics[width = 3.0in]{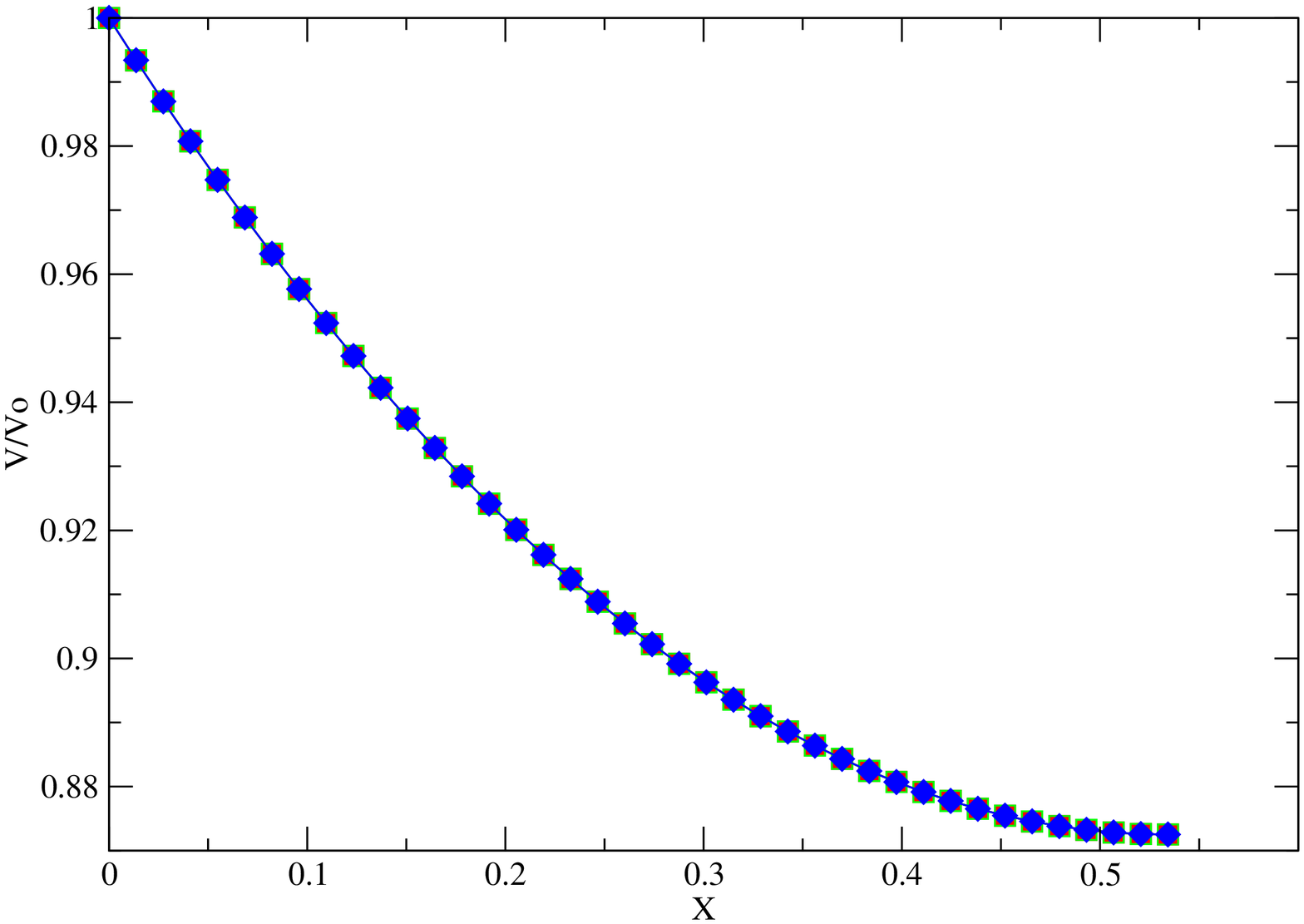} 
} 
\end{center}
\caption{\textbf{Comparison between compact,analytical and central  difference schemes ( point soma), $N=10,20,30,40$,{ \color{red}$\blacksquare$}:compact;{\color{green}$\blacksquare$}:analytical;{\color{blue}$\blacksquare$}:central}}
\label{fig:cabunpanumcent}
\end{figure}

$\frac{V}{V_{o}}$ is plotted against $X$ for the compact difference scheme,central difference scheme and the analytical solution for sealed end and results are shown in (Fig.~\ref{fig:cabunpanumcent}). The second derivative $\frac{\partial^{2}V}{\partial x^{2}}$ is approximated using equation $4$. By choosing coefficients as given below, it gives fourth order accuracy. The coefficients are (~\cite{tuckwell1:88a}, equation $2.2.6$):
 \begin{displaymath}
  \beta = 0,\hspace{2mm} c = 0,\hspace{2mm} a = \frac{4(1-\alpha)}{3},\hspace{2mm} b = \frac{1(-1 +10\alpha)}{3},\hspace{2mm} \alpha = \frac{1}{10},\hspace{2mm} \alpha1 = \frac{1}{10}
 \end{displaymath}
 The equation for the fourth order central difference scheme is calculated using (~\cite{mathews:04}, Table $6.4$):
\begin{equation}
 V''(i) =\frac{ [\frac{-1}{12}V(i-2)+\frac{4}{3}V(i-1)-\frac{5}{2}V(i) + \frac{4}{3}V(i+1)-\frac{-1}{12}V(i+2)]}{h^2}  , 2 \leq i \leq N-1
\end{equation}  
                         
The boundaries for the central difference scheme are calculated using :
\begin{equation}
V''(1) = (\frac{4V(1+1)-3V(1)-V(1+2)}{2h})
\end{equation}
\begin{equation}
V''(N) = (\frac{4V(N-1)-3V(N)-V(N-2)}{2h}) 
\end{equation} 
The points next to the boundary are calculated using :
\begin{equation}
 V''(2)= \frac{\frac{15}{4}V(2)-\frac{77}{6}V(N-2)+\frac{107}{6}V(4)-13V(5)+\frac{61}{12}V(6)-\frac{5}{6}V(7)}{h^2}
\end{equation} 
\begin{equation}
 V''(N-1)= \frac{\frac{15}{4}V(N-1)-\frac{77}{6}V(N-2)+\frac{107}{6}V(N-3)-13V(N-4)+\frac{61}{12}V(N-5)-\frac{5}{6}V(N-6)}{h^2}
\end{equation} 
where $h = \Delta x$. 
\subsubsection*{Comparison of error between compact and central difference schemes, $N=10,20,30,40$}
\begin{figure}[!ht]
\begin{center}
\psfrag{V/Vo}{$V/Vo$}
\psfrag{X(lambda)}{$X(lambda)$}
\psfrag{a}{$a$}
\psfrag{b}{$b$}
\psfrag{c}{$c$}
\psfrag{d}{$d$}
\subfigure[{$N=10$}]{
\includegraphics[width = 3.0in]{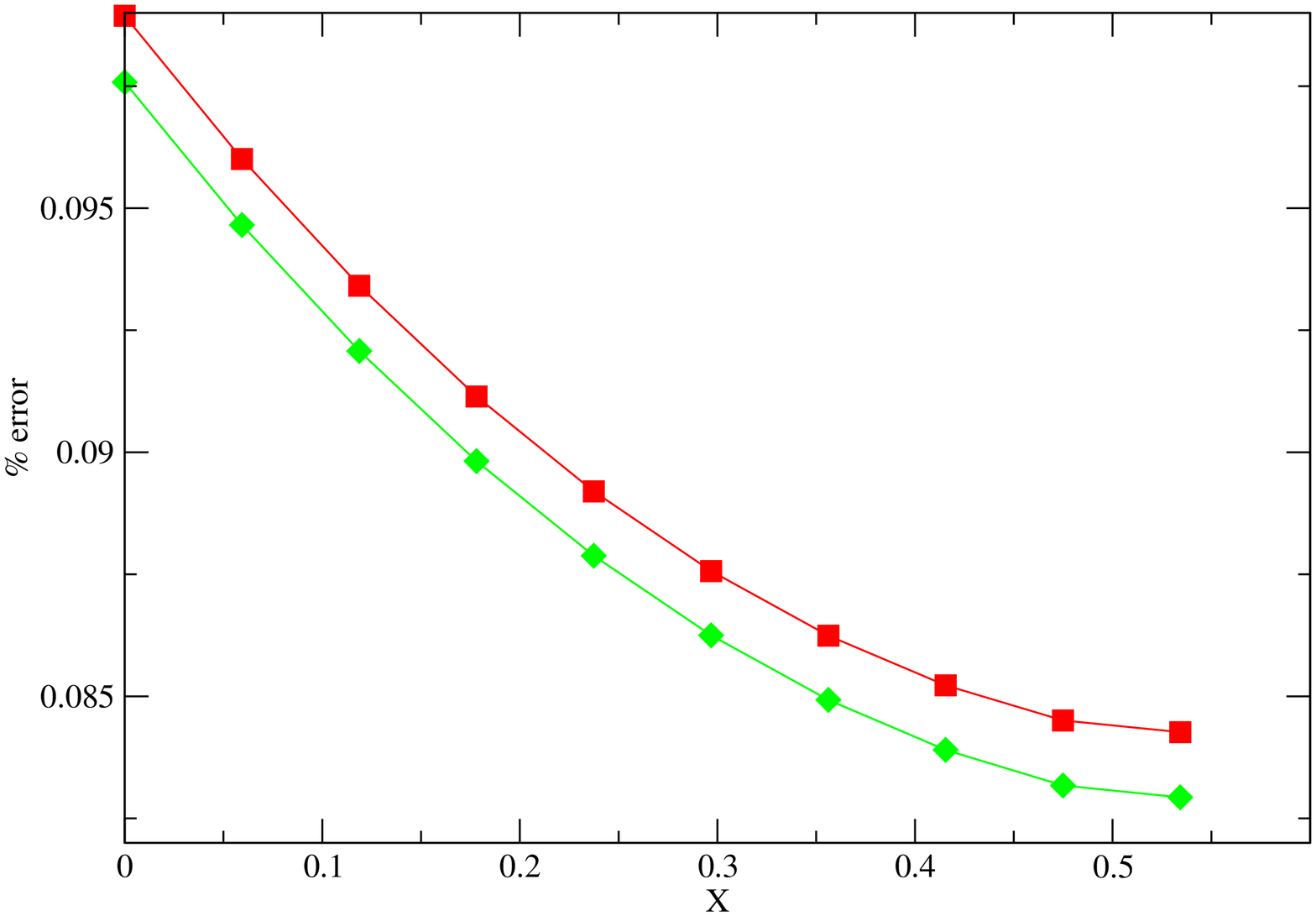}
}
\subfigure[{$N=20$}]{
\includegraphics[width = 3.0in]{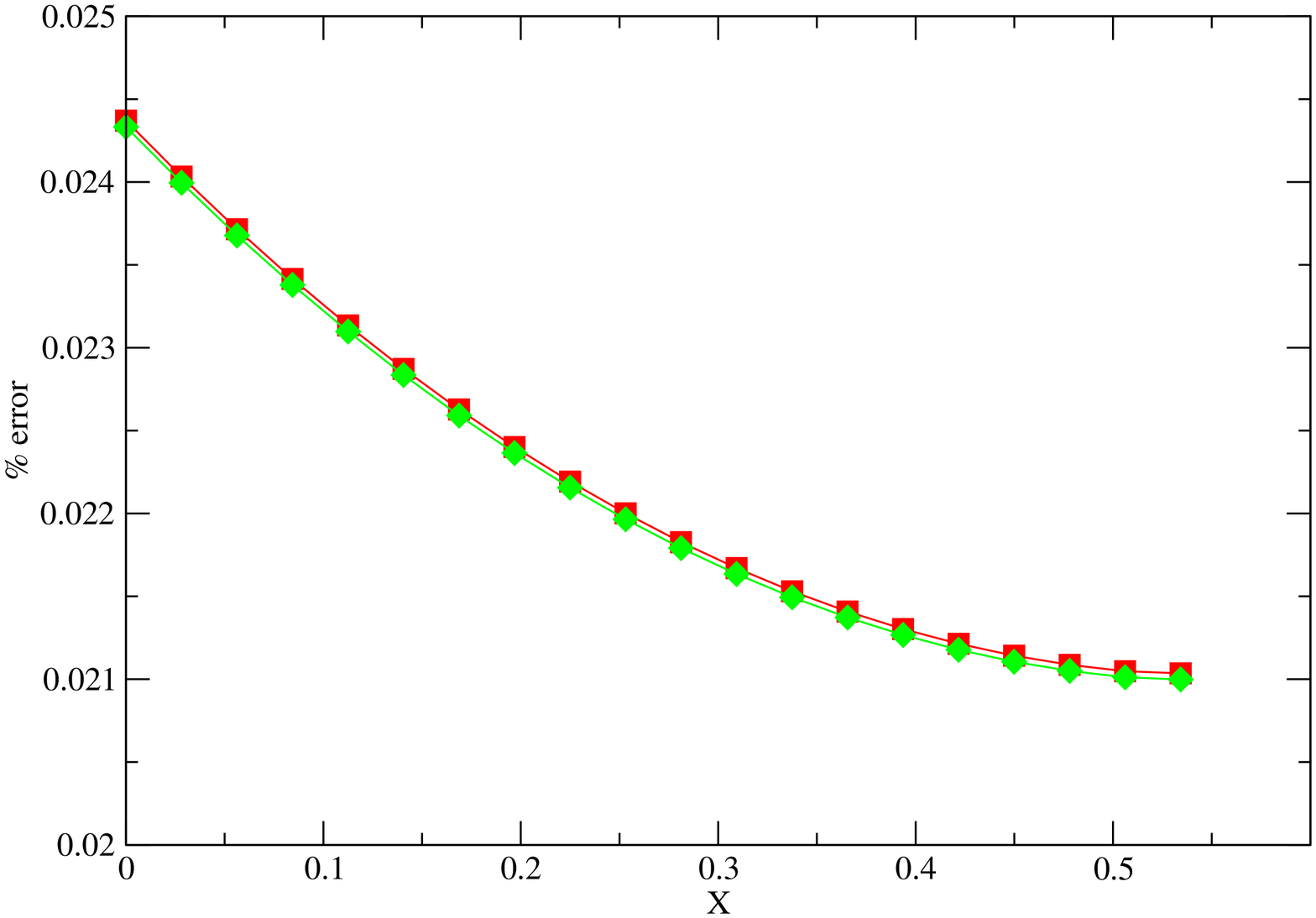}
}
\subfigure[{$N=30$}]{
\includegraphics[width = 3.0in]{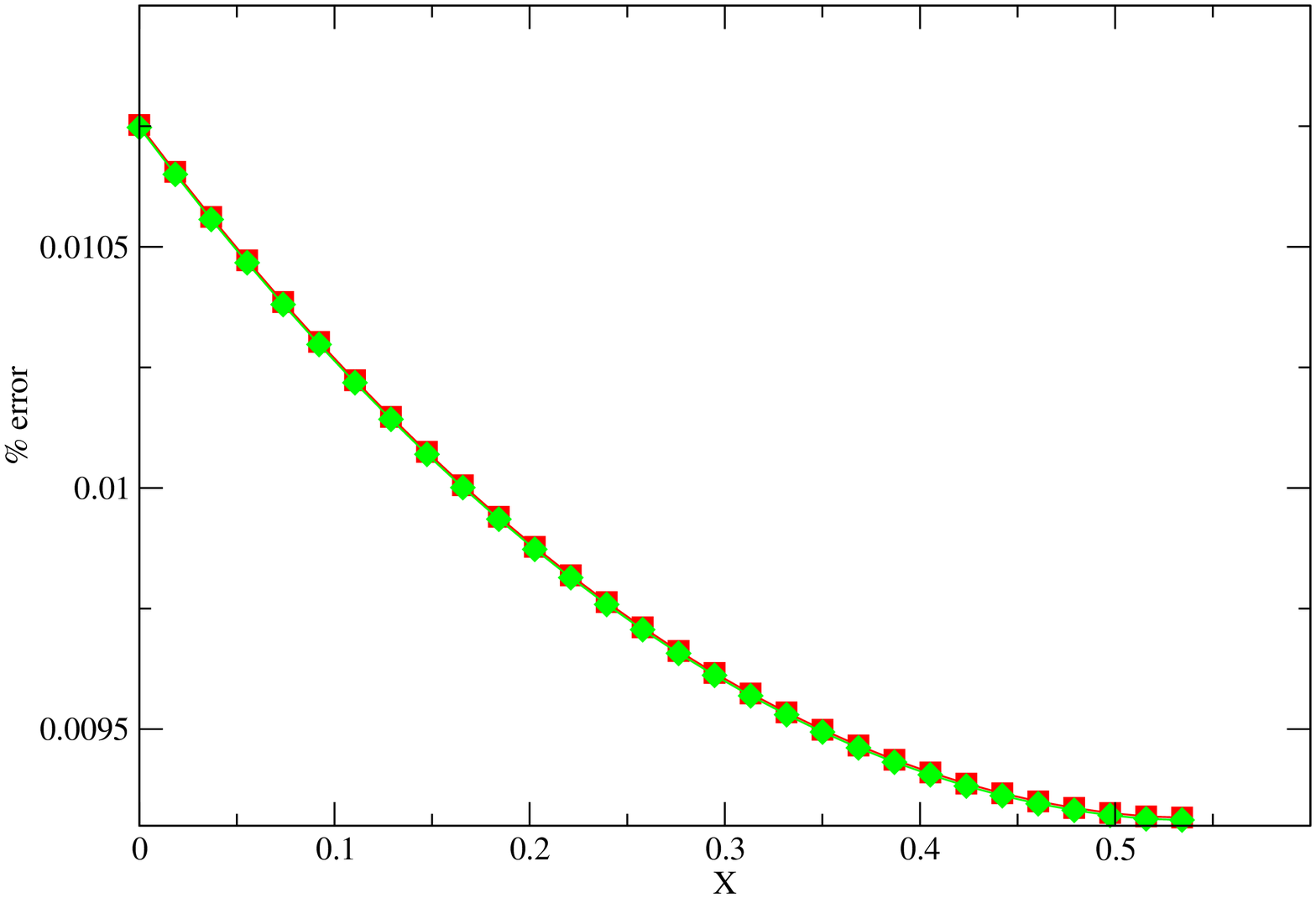}
}
\subfigure[{$N=40$}]{
\includegraphics[width = 3.0in]{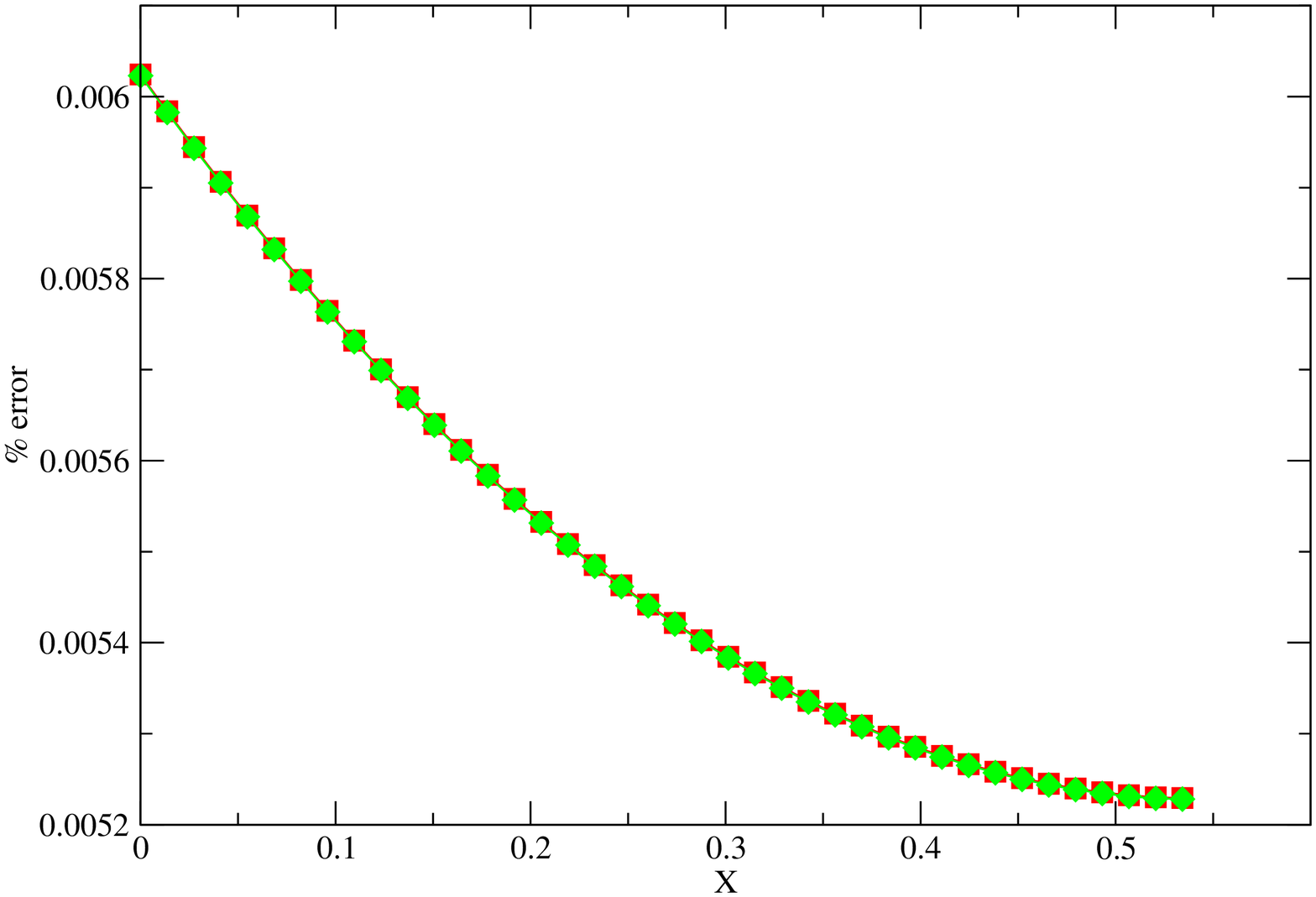}
}
\end{center}
\caption{\textbf{Comparison of error between compact and central  difference  schemes(point soma), $N = 10,20,30,40 $,{ \color{red}$\blacksquare$}:compact;{\color{green}$\blacksquare$}:central}}
\label{fig:cabunpanumcenterror}  
\end{figure}

\begin{equation}
\%Error1 = 100*|\frac{V_{comp}-V_{anal}}{V_{anal}(1)}|
\end{equation} 
\begin{equation}
\%Error2 = 100*|\frac{V_{cent}-V_{anal}}{V_{anal}(1)}|
\end{equation} 
$\%Error 1$ and $\%Error 2$ are plotted against  $X$. The results are shown in (Fig.~\ref{fig:cabunpanumcenterror}).
\subsubsection*{Dependence of voltage changes on dendritic diameter }
\begin{figure}[!ht]
\begin{center}
\psfrag{V/Vo}{$V/Vo$}
\psfrag{X(lambda)}{$X(lambda)$}
\subfigure[{sealed end}]{
\includegraphics[width = 3.0in]{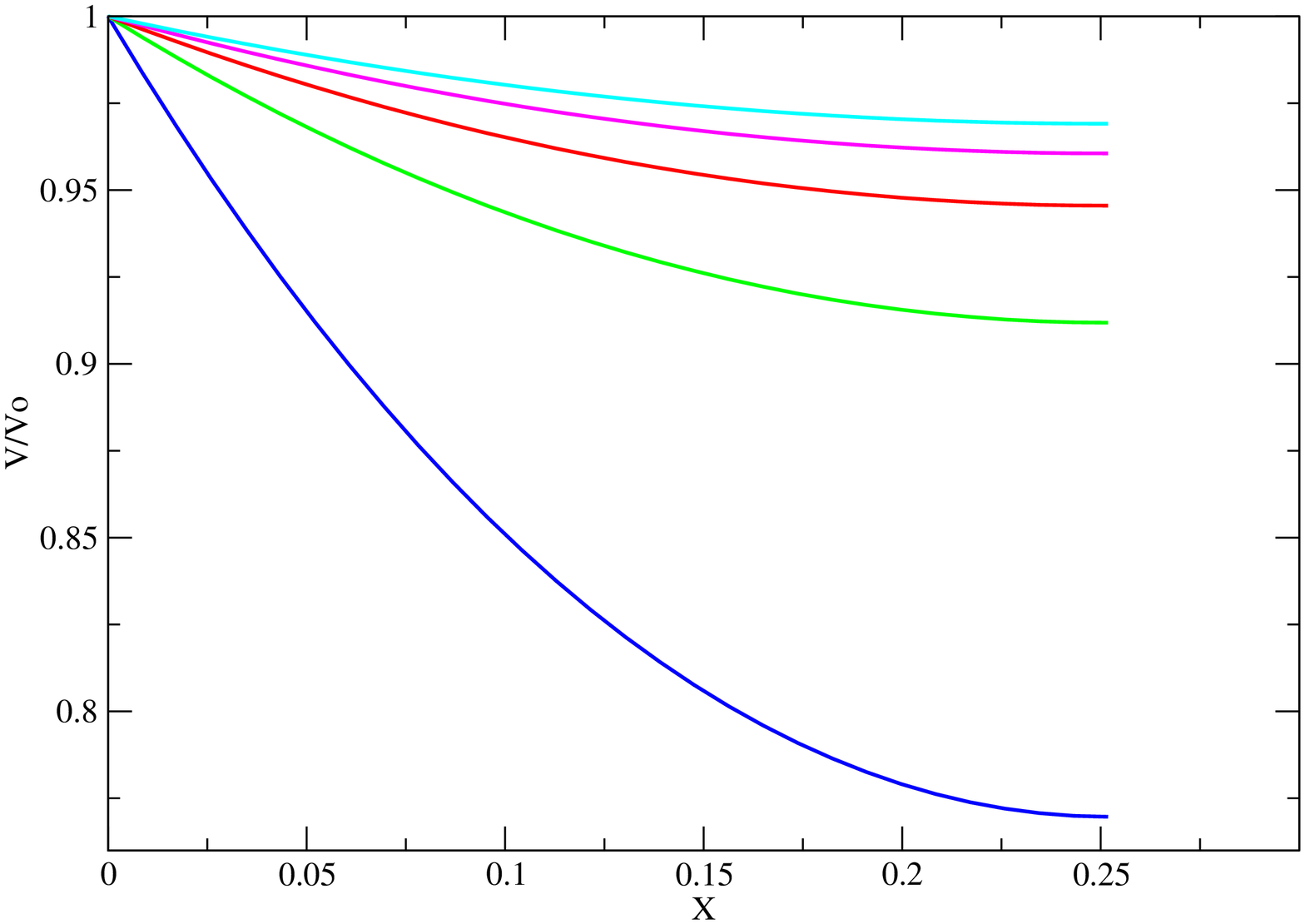}
 }
\subfigure[{killed end}]{
\includegraphics[width = 3.0in]{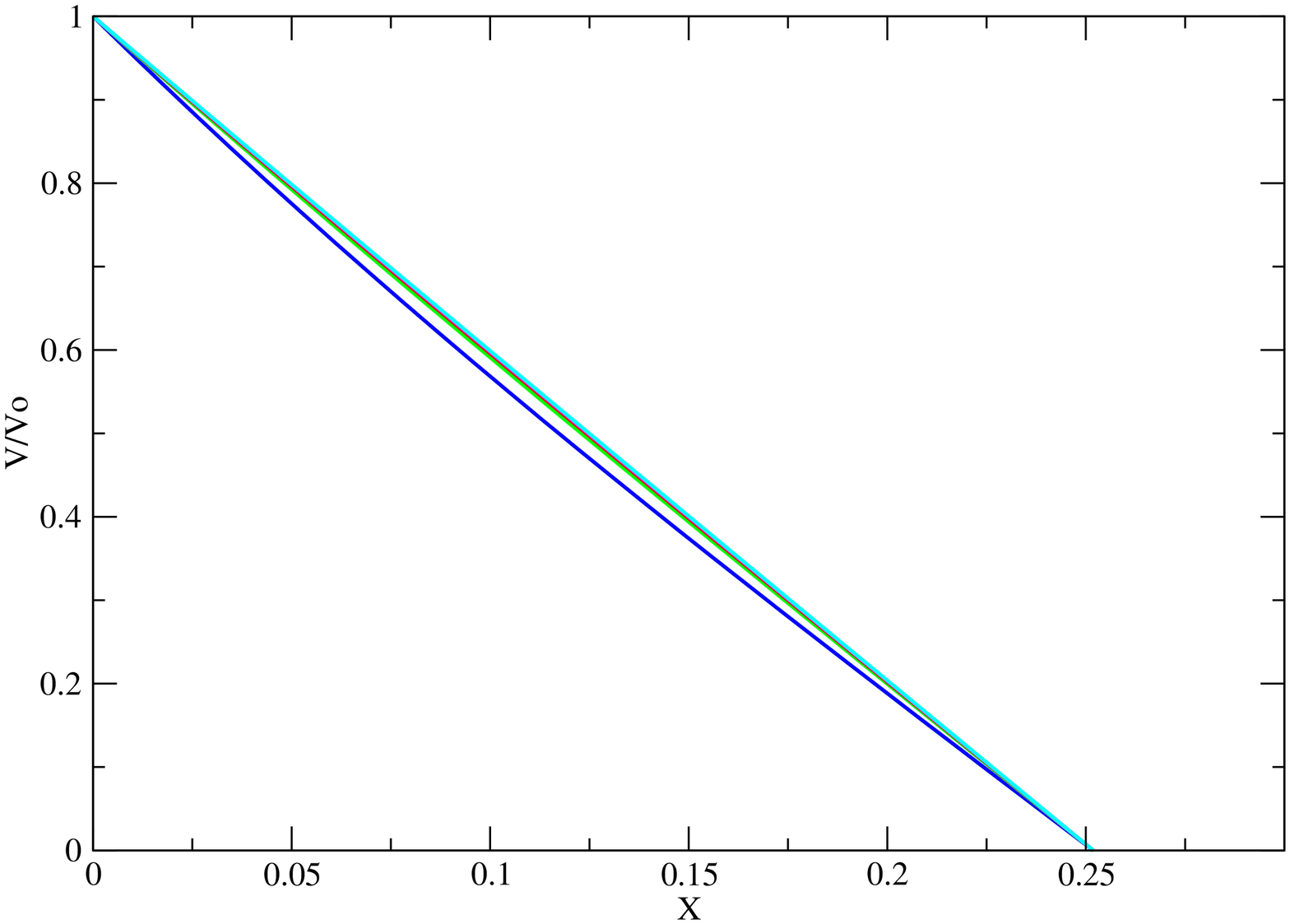}
}
\end{center}
\caption{\textbf{Effect of diameter, sealed and killed end,(point soma), {\color{blue} ------}: $d=1.85*10^-4$(cm); {\color{green} ------}: $d=3*1.85*10^-4$;{\color{red}----}:$d=5*1.85*10^-4$;{\color{magenta}----}:$d=7*1.85*10^-4$;{\color{cyan}----}:$d=9*1.85*10^-4$}}
\label{fig:cabunpa2sedendia}
\end{figure}

Current is injected at $i=1$ for dendrites of various diameter and $\frac{V}{V_{0}}$ is plotted versus $X$. The results are given in (Fig.~\ref{fig:cabunpa2sedendia}). 
\subsection*{Point soma dendrite, current injection as $x=l$}
\subsubsection*{Current injection at $i=N$, sealed and killed end, Evolution of voltage }
\begin{figure}[!ht]
\begin{center}
\psfrag{V in mV}{$V in mV$}
\psfrag{X(\lambda)}{$X(\lambda)$}
\subfigure[{sealed end}]{
\includegraphics[width = 3.0in]{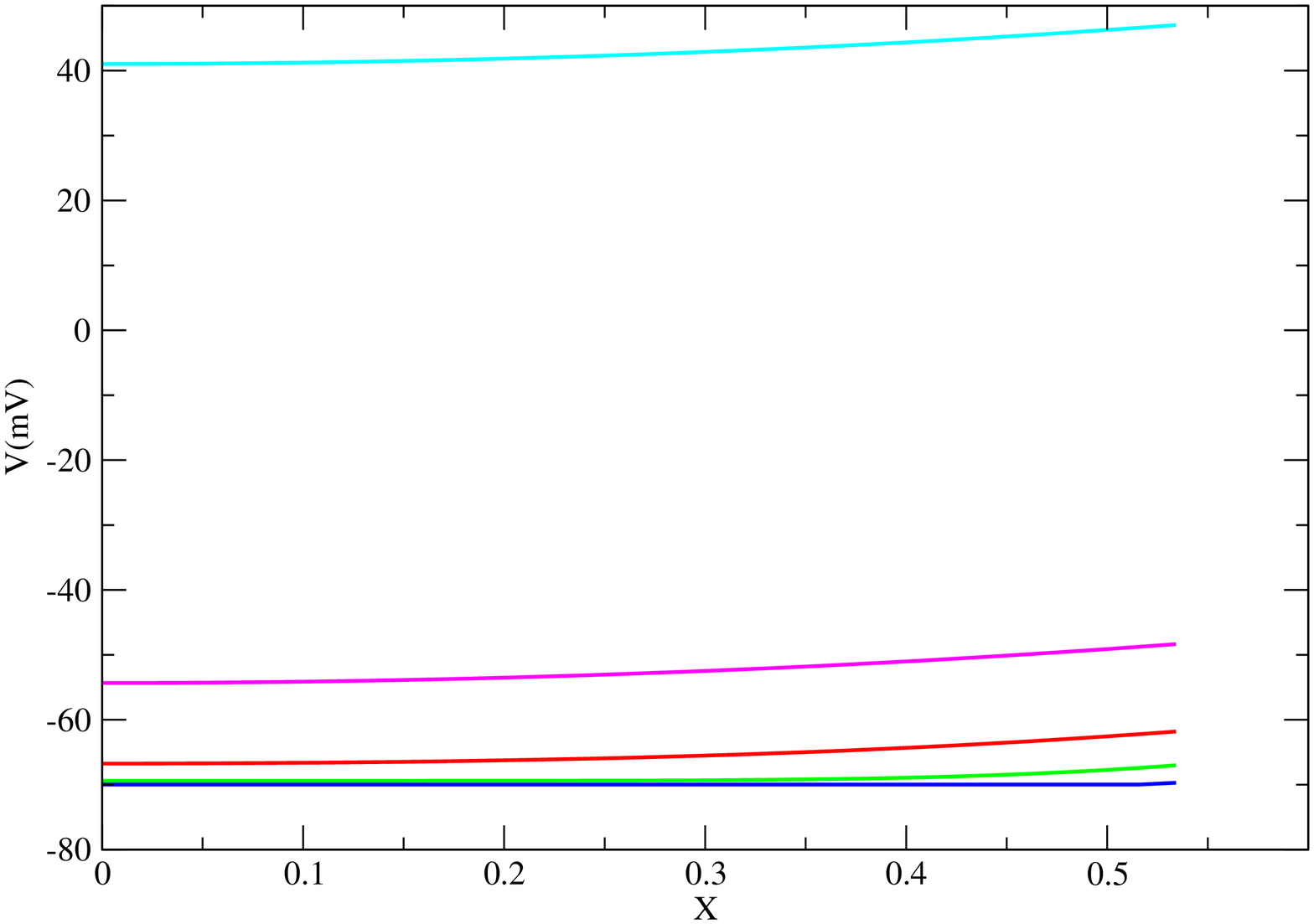}
}
\subfigure[{killed end}]{
\includegraphics[width = 3.0in]{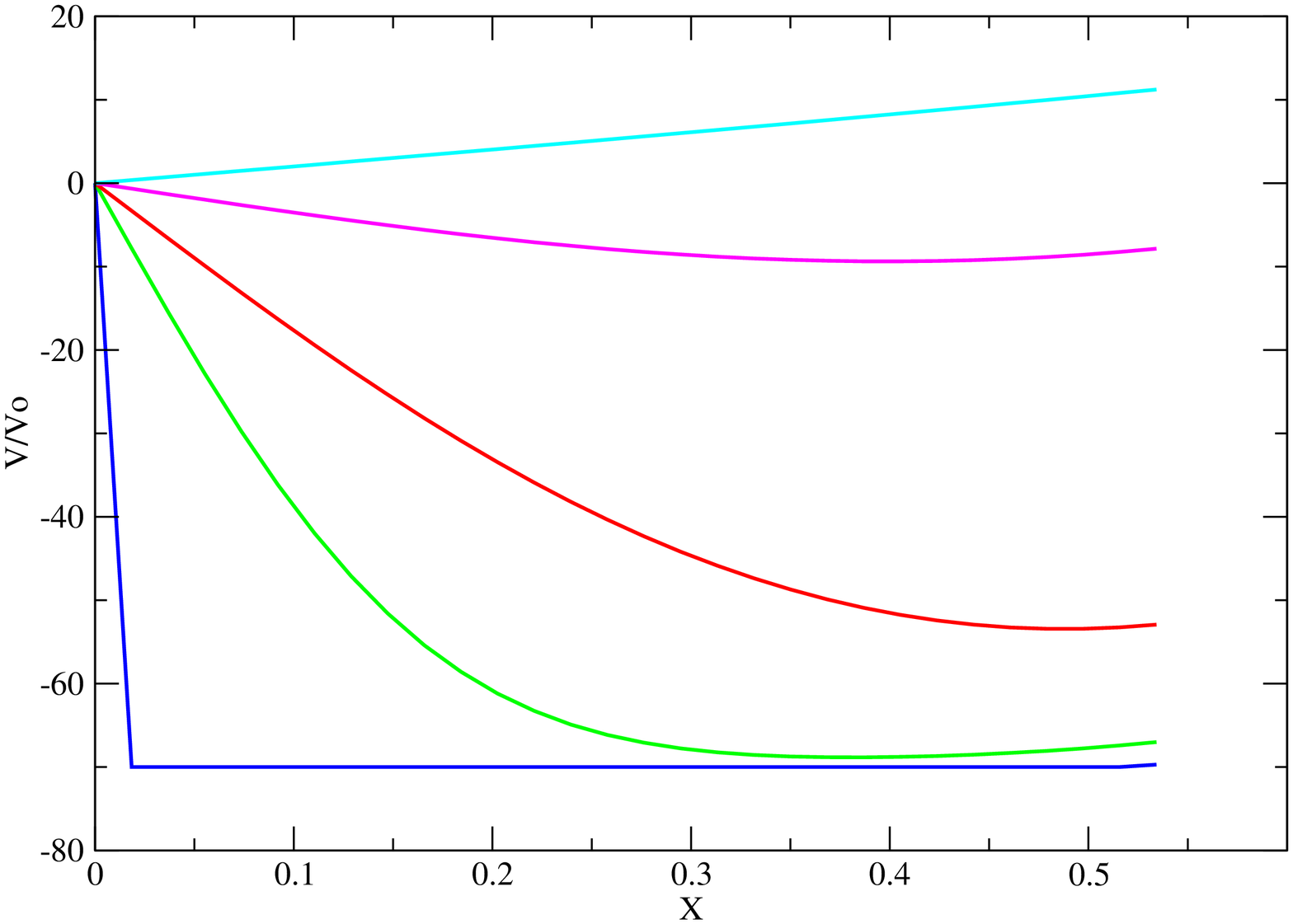}
}
\end{center}
\caption{\textbf{Evolution of voltage,current injected at $i= l$(point soma), sealed end and killed end boundaries,{\color{blue} ------}: $n=1$; {\color{green} ------}: $n=100$;
{\color{red}-----}:$n==500$;{\color{magenta}-----}:$n==2000$;{\color{cyan}----}:$n==294670$} } 
\label{fig:cabunpainjl}
\end{figure}

The current is injected at the end opposite to the soma and the voltage evolution from $t = 0.0016969$msec to $t= 500$msec is plotted. The intial condition is shown in equation $8$. The boundary condition at $i=1$ is chosen to be either sealed end given by equation $11$ or killed end, equation $13$.The boundary condition at $i=N$ is the current injection boundary condition shown by equation $10$. Results are shown in (Fig.~\ref{fig:cabunpainjl}).
\subsection*{Point soma dendrite, current injection as $x=\frac{l}{2}$}
The current is injected at $i=\frac{N}{2}$. The initial condition is given by equation $8$. The boundary conditions at $i=1$ and $i=N$ can either be sealed end given by equations $11,12$ or killed end given by equations $13,14$.
\subsection*{Lumped soma, current injection at soma }
\subsubsection*{Current injection at soma, lumped soma boundary condition,Evolution of voltage}
Here the current is injected at the soma as shown in (Fig.~\ref{fig:pointsomadendx0}) and the evolution of voltage at various times is observed.\\
\begin{equation}
 t= (\Delta T)(n)(\tau) 
\end{equation}
For $N=30$,$\tau = 20$ msec,the voltage evolution is looked  at  $t$ as shown in (Table~\ref{tab:niter}). The results for both sealed and killed end boundary conditions at $i=N$ can be seen in (Fig~\ref{fig:sevoltNoct2}). 
\section*{Discussion}
It might be apt here to quote Wilfred Rall ~\cite{rall:08},' An important basic principle of prudent research, both theoretical and experimental, is not to tackle too many complications at once. This was my reason for beginning my modeling with uniform, passive membrane, and with idealized geometry. Once you solve the reduced problem, you can then begin to deal with some of the complications judged to be functionally important. ' Keeping this advice in mind, we have demonstrated here the usage of the compact difference scheme to a passive soma dendrite construct under varying experimental and morphological situations.\par\parindent = 1 in Our results show that the scheme is robust under all the conditions and shows a good fit with the analytical solution even at  $N = 10$. By looking at the percent error $EP1$ ( point soma),(Table.~\ref{tab:pointerror}) at $N=10$, it can be seen that it is almost the same for compact and central schemes. The error due to the central scheme is about $0.001$ less than that due to the compact scheme. \par\parindent = 1 in Lele~\cite{lele:92}defines resolving efficiency as a difference between the modified wavenumbers of spectral scheme and the differencing scheme. This is specific to a scheme for any given error. Thus it can be seen that at error $ \epsilon <= 0.001$ , the resolving efficiency of the fourth order compact scheme is $0.22$ which is greater than that of the fourth order central scheme which is $ 0.17$. See (Table.~\ref{tab:resolving}). It is also seen that the resolving efficiency of the sixth order compact scheme is 0.38.\par\parindent = 1in
 The simulations dealing with evolution of voltage show that when current is injected at $i=1$, the voltage evolves to a higher value for both sealed and killed end. The cable reaches a steady state  faster for the killed end than the sealed end. This is to be expected as there is no leakage at the sealed boundary. In the lumped soma case, the voltage reached at steady state is less than that at the point soma. This could be due to  the R-C circuit comprising the lumped soma which results in some current injected going to charge this resulting in a smaller value of current flowing through the cable. The shape of the voltage evolution when current is injected at $i=N$(point soma) is different for both sealed and killed end and it also evolves to a lesser voltage at steady state than current injection at $i=1$. The rate of evolution is also much slower in this case. Finally, when current is injected at $i=\frac{N}{2}$, the resulting plots show predictable curves expected when either both ends are sealed or killed. \par\parindent = 1in As $\lambda $ is directly proportional to the square root of the diameter of the dendrite,one can see that as the diameter increases (Fig~\ref{fig:cabunpa2sedendia}), the lambda increases too and the plots for both sealed and killed end boundaries give expected results. See (Table.~\ref{tab:Lambda})\par\parindent= 1in The numerical integration under consideration is an explicit scheme. In compact schemes, explicit schemes can only be used. This is unlike central schemes where both explicit and implicit techniques can be used. In the compact scheme used  here, the stability is linked to $\Delta T$ or the time step that is being used. Various values of $\Delta T$ used are shown in (Table~\ref{tab:DeltaT}). It is observed that this results in slower computations. \par\parindent= 1in In the simple cable chosen here,  it is not possible to see the spatial resolving efficiency of the compact scheme. It is expected that the spatial resolving power of the compact scheme can be viewed with greater complexity of the model and also by choosing schemes with different coefficiants than the sixth order tridiagonal scheme chosen here. From Lele~\cite{lele:92} it can be seen that an eighth order tridiagonal scheme can yield at $ \epsilon <= 0.001$, a resolving efficiency of 0.48 and a spectral like pentadiagonal scheme can yield a resolving efficiency of 0.84. \par\parindent=1in In this paper, it has been shown that the compact scheme can be used to solve the cable equation under various morphological and experimental conditions. Ongoing work looks at the use of this tool in the case of branched,tapered and active dendrites where the complex geometry could yield situations where the compact  scheme is more efficient.


\section*{Acknowledgments}
AG would like to thank Joseph Mathew, Professor, Department of Aerospace Engineering, Indian Institute of Science, Bangalore for suggesting the use of the compact difference scheme as an alternative to spectral methods in the numerical analysis of the cable equation. He also taught AG the scheme and has smoothened many problems during its implementation.AG would like to acknowledge the support provided by A.K Gupta( currently NIMHANS, Bangalore), M.D Nair and K.Radhakrishnan of Sree Chitra Tirunal Institute of Medical Sciences and Technology, Tiruvananthapuram.  AG would also like to acknowledge the support of Joseph Mathew, G. Rangarajan, V.Nanjundiah and P. Balaram in making arrangements to work at the Indian Institute of Science. AG also acknowledges the support provided by K.R Srivathsan,faculty and staff of the Indian Institute of Information Technology and Management, Kerala for making arrangements to work there. AG thanks Maya Ramachandran and Venugopalan for acquiring necessary references from the library of the National University of Singapore. Thanks are due to Vimal Joseph,Ganesh, Sujit of SPACE, Tiruvananthapuram, Shiv Chand of IIITM-K, Rajdeep Singh, Rani and other staff of SERC, IISc for help with computer hardware and software. 

\bibliography{natur}

\clearpage
\section*{Figure}
\begin{figure}[!ht]
\begin{center}
\psfrag{soma}{$soma$}
\psfrag{dendrite}{$dendrite$}
\psfrag{Rs}{$Rs$}
\psfrag{Cs}{$Cs$}
\psfrag{Io}{$Io$}
\psfrag{Id}{$Id$}
\psfrag{Is}{$Is$}
\psfrag{X=0}{$X=0$}
\psfrag{X=L}{$X=L$}
\includegraphics[width = 3.27in]{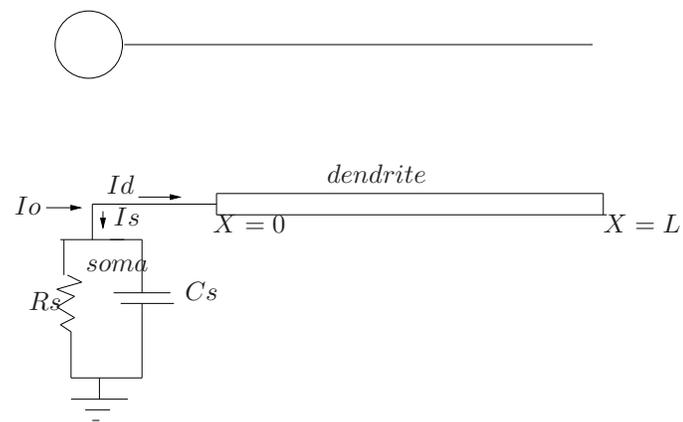}
\end{center}
\caption{\textbf{Dendrite with lumped soma}}
\label{fig:lumpedsoma}
\end{figure}

\clearpage

\begin{figure}[!ht]
\begin{center}
\subfigure[{sealed end}]{
\includegraphics[width = 3.0in]{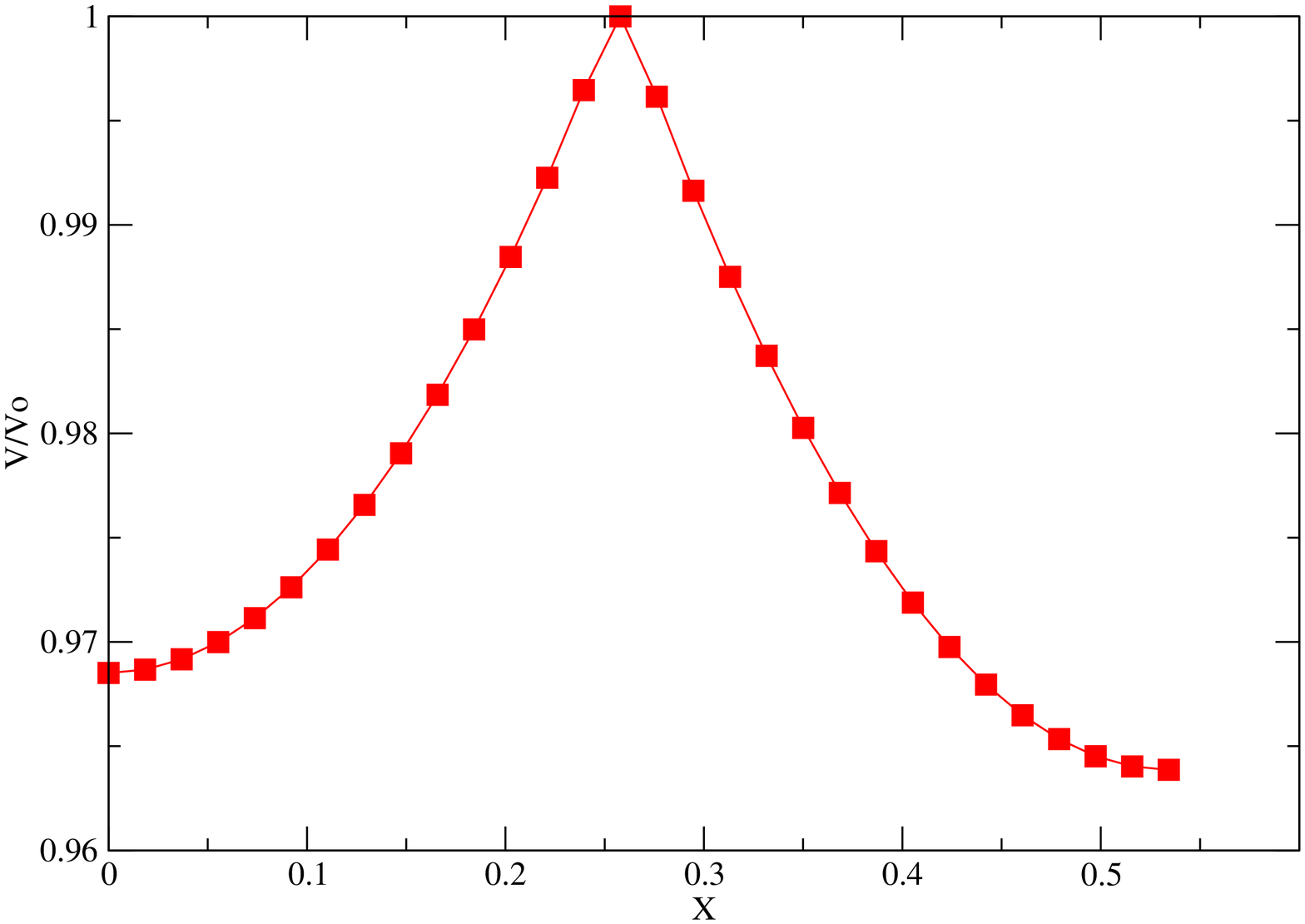}
}
\subfigure[{killed end}]{
\includegraphics[width = 3.0in]{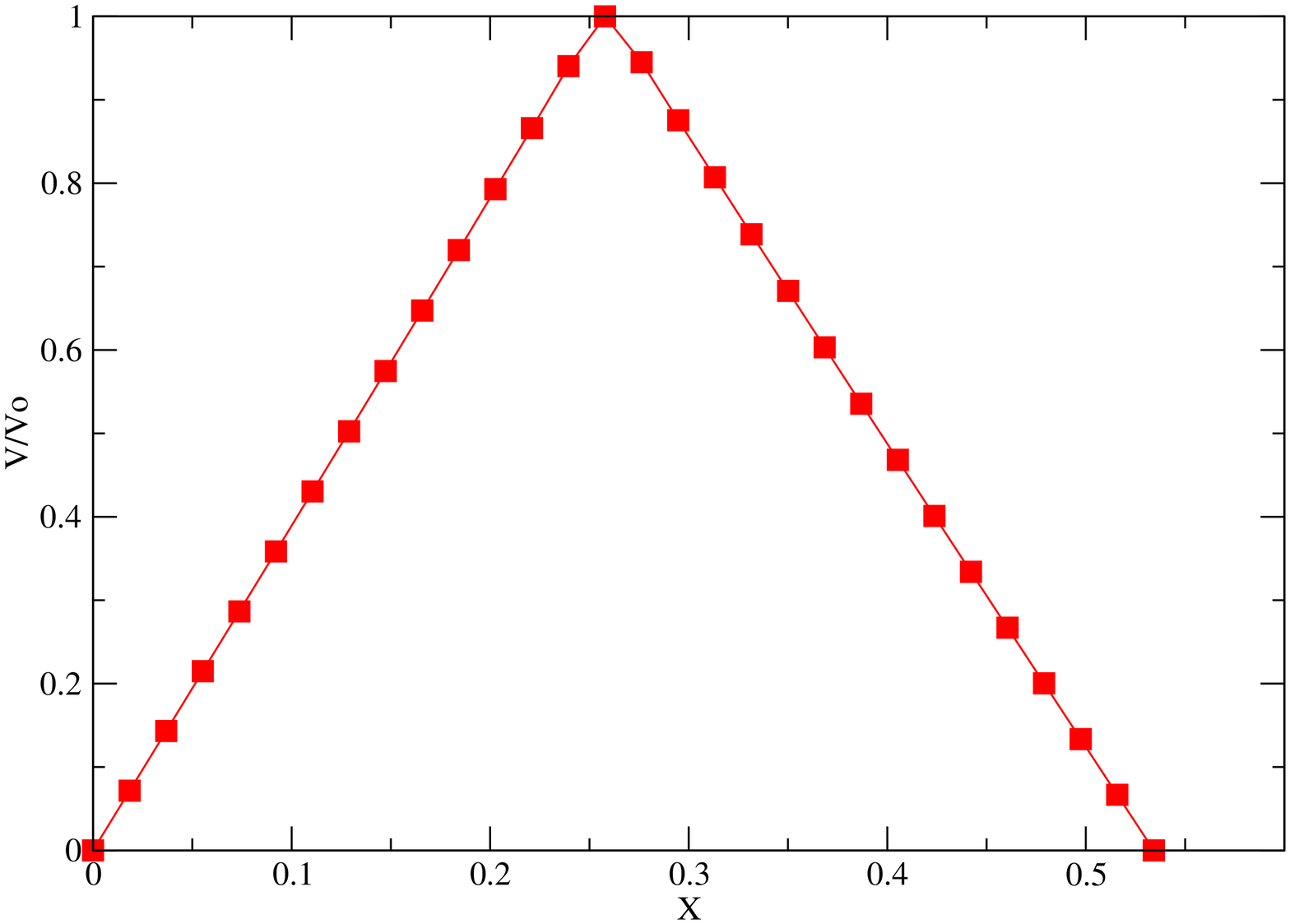}
}
\end{center}
\caption{\textbf{Current injected at $i= N/2$,(point soma) sealed end and killed end boundaries{\color{red}$\blacksquare$}:compact}}
\label{fig:cabunpainjN2}
\end{figure}

\clearpage

\section*{Tables}
\begin{table}[!ht]
\caption{\bf{Values of $\Delta T$ at various $N$}}
\centering
\begin{tabular}{c rr}
\hline 
$N$& \multicolumn{2}{c}{$\Delta T$} \\
\hline
$10$ & $8.8088e-04$\\
$20$& $1.9765e-04$ \\
$30$ & $8.4841e-05$  \\
$40$ & $4.6911e-05$ \\
$80$ & $1.1433e-05$\\
\hline
\end{tabular} 
\label{tab:DeltaT}
\end{table}
\begin{table}[!ht]
\caption{\bf{Values of $t$ in msec at various $n$}}
\centering
\begin{tabular}{c rr}
\hline 
$n$& \multicolumn{2}{c}{$t$} \\
\hline
$1$ & $0.0016968$\\
$100$& $0.16968$ \\
$500$ & $0.84840$  \\
$2000$ & $3.3936$ \\
$294670$ & $500$\\
\hline
\end{tabular} 
\label{tab:niter}
\end{table}
\begin{table}[!ht]
\caption{\bf{Parameters of dendrite used in simulation}}
\centering
\begin{tabular}{c rr}
\hline 
Parameter& \multicolumn{2}{c}{Values} \\
\hline
length & $400 \mu$\\
diameter& $2*1.85 \mu$ \\
$R_{m}$ & $20*10^3 \Omega.cm^2 $  \\
$R_{i}$ & $330  \Omega.cm $ \\
$C_{m}$ & $1 \mu farad/cm^2 $\\
$\tau$ & $20 $ msec \\
$Iinj$ & $ 0.1 $nanoamperes \\
\hline
\end{tabular} 
\label{tab:parameters}
\end{table}
\oddsidemargin -2.0cm
\evensidemargin -2.0cm
\begin{table}[ht]
\caption{Comparison of percentage error ($E_{P}$) between compact and central for $N=10$(point soma) }
\centering
\begin{tabular}{c c c c c c c c c c c }
\hline\hline
Error & $i=1$ & $i=2$ & $i=3$ & $i=4$ & $i=5$ & $i=6$ & $i=7$ & $ i=8$ & $ i=9$ & $ i=10$ \\[0.5ex]
\hline
Compact & 0.098941& 0.096008 & 0.093412 & 0.091144 & 0.089198 & 0.087567 & 0.086244 & 0.085225 & 0.084506 & 0.084267\\
Central & 0.097582 & 0.094655 & 0.092074 & 0.089819 & 0.087880 & 0.086251 & 0.084926 & 0.083901 & 0.083172 & 0.082929 \\
\hline 
\end{tabular}
\label{tab:pointerror}
\end{table}
\oddsidemargin -2.0cm
\evensidemargin -2.0cm
\begin{table}[h]
\caption{Resolving Efficiency $\epsilon$ of the second derivative schemes,(~\cite{lele:92}, Table $5$)}
\centering
\begin{tabular}{c c c c }
\hline\hline
Scheme & $\epsilon = 0.1$ & $\epsilon = 0.01$ & $\epsilon = 0.001$  \\[0.5ex]
\hline
Fourth order central & 0.59& 0.31 & 0.17\\
Fourth order compact  & 0.68 & 0.39 & 0.22 \\
Sixth order tridiagonal & 0.80 & 0.55 & 0.38 \\
\hline 
\end{tabular}
\label{tab:resolving}
\end{table}
\begin{table}[!ht]
\caption{\bf{Values of $\lambda$ (cm) at various $d$(cm) }}
\centering
\begin{tabular}{c rr}
\hline 
$d$& \multicolumn{2}{c}{$\lambda$} \\
\hline
$1.85*10^-4$ & $0.052223$\\
$3*1.85*10^-4$& $0.091701$ \\
$5*1.85*10^-4$ & $0.11839$  \\
$7*1.85*10^-4$ & $0.14008$ \\
$9*1.85*10^-4$ & $0.15883$\\
\hline
\end{tabular} 
\label{tab:Lambda}
\end{table}
\end{document}